\documentclass[a4paper,11pt]{article}
\usepackage{float}
\usepackage[pdftex]{graphicx}
\usepackage{subcaption}
\usepackage[T1]{fontenc}
\usepackage{lmodern}
\usepackage{slashed}
\usepackage[utf8]{inputenc}
\usepackage[english]{babel}
\usepackage{microtype}
\usepackage{cite}
\usepackage{amsmath,amssymb,amsfonts,amsthm}
\usepackage{mathtools,mathrsfs,calligra,aurical}
\usepackage[nottoc,notlot,notlof]{tocbibind}
\usepackage{upgreek}
\usepackage{mathtools}
\numberwithin{equation}{section}
\allowdisplaybreaks
\usepackage[all]{xy}
\usepackage{xcolor}
\usepackage{graphicx}
\graphicspath{{images/}}
\usepackage{geometry}
\usepackage[toc,page]{appendix}
\usepackage{hyperref}
\usepackage[normalem]{ulem}

\usepackage{bm}
\usepackage{ragged2e}
\usepackage{appendix}
\usepackage{slashed}
\usepackage{bbold}
\usepackage{cancel}

\definecolor{blue-violet}{rgb}{0.54, 0.17, 0.89}
\definecolor{PineGreen}{cmyk}{0.92, 0, 0.59, 0.25}
\definecolor{YellowOrange}{cmyk}{0, 0.42, 1, 0}
\definecolor{orange}{rgb}{0.95, 0.5, 0.1}

\newcommand{\be}{\begin{equation}}
\newcommand{\bea}{\begin{eqnarray}}

\newcommand{\ee}{\end{equation}}
\newcommand{\eea}{\end{eqnarray}}

\def\d{\partial}
\def\a{\alpha}

\DeclareMathAlphabet{\mathpzc}{OT1}{pzc}{m}{it}

\begin{document}

\begin{titlepage}
\begin{flushright}
\par\end{flushright}
\vskip 0.5cm
\begin{center}
\textbf{\LARGE \bf  Supersymmetric AdS Solitons and the interconnection of different vacua of ${\cal N}=4$ Super Yang-Mills}\\
\vskip 5mm

\vskip 1cm

\large {\bf Andr\'{e}s Anabal\'{o}n}$^{~a ~b}$\footnote{anabalo@gmail.com}, \large {\bf Horatiu Nastase}$^{~b}$\footnote{horatiu.nastase@unesp.br} and \large {\bf Marcelo Oyarzo}$^{~a}$\footnote{moyarzoca1@gmail.com}

\vskip .5cm 

$^{(a)}${\textit{Departamento de F\'isica, Universidad de Concepci\'on, Casilla, 160-C, Concepci\'on, Chile.}}\\ \vskip .1cm
$^{(b)}${\textit{Instituto de F\'isica Te\'orica, UNESP-Universidade Estadual Paulista \\
Rua Dr. Bento T. Ferraz 271, Bl. II, Sao Paulo 01140-070, SP, Brazil.}}
\end{center}
\begin{abstract}
{We find AdS soliton solutions in 5-dimensional gauged supergravity, obtained from the $S^5$ compactification of type IIB, 
with a dilaton saturating the Breitenlohner-Freedman bound. The solutions depend on the value of the periodicity of an $S^1$ cycle 
and the boundary values for two $U(1)$ gauge fields, and give a scalar VEV in the dual field theory. 
At certain values of the gauge sources we have supersymmetric solutions, 
corresponding to supersymmetric flows, which are a deformation 
of the Coulomb Branch flow in ${\cal N}=4$ SYM. 
The solutions parameterize quantum phase transitions between 
a discrete spectrum phase, a continuous above a mass gap phase, and a continuous without a mass gap
phase, in 2+1 dimensions. We analyze the phase diagram 
in terms of the QFT sources and we find that for every value for them, there are always two branches of supergravity solutions. 
We find that these two branches of solitons correspond to two possible vacua existing in the dual QFT when fermions 
are anti-periodic on an $S^1$. We describe the interconnection of these states in the QFT at strong 't Hooft coupling 
in the large $N$ limit. In 10 dimensions, our solutions are related to 
deformations of D3-brane distributions.}
\end{abstract}

\vfill{}
\vspace{1.5cm}
\end{titlepage}

\setcounter{footnote}{0}
\tableofcontents

\section{Introduction}

In the AdS/CFT correspondence, the Witten model \cite{Witten:1998zw}, 
corresponding to a scaling $M\rightarrow \infty$ of a Schwarzschild-AdS
black hole, or to a near-horizon near-extremal limit of D3-branes, is interpreted as dual to ${\cal N}=4$ SYM at finite temperature or, 
after a double Wick rotation by replacing the periodic time $t$ with a Kaluza-Klein (\textbf{KK}) angular coordinate $\phi$, and a reduction on $\phi$, 
as dual to 3-dimensional pure glue theory ($\equiv QCD_3$; fermions are antiperiodic, so massive and scalars gain a mass
at one-loop, from the fermions), coupled to extra modes at the KK scale $T_{\rm KK}=1/R_\phi$, 
and one obtains a discrete spectrum of states. 
This is also similar to what one obtains by cutting off AdS space in the IR (the ``hard-wall'' model). 

But there are other behaviours possible from deforming ${\cal N}=4$ SYM. One such is the ``Coulomb Branch (CB)'' 
deformation of ${\cal N}=4$ SYM, by a scalar operator of dimension $\Delta=2$, studied in \cite{Freedman:1999gk}. 
One obtains two possible metrics, described by dimensionless parameters $\pm \ell^2/L^2$, describing either discrete
states (for minus sign) or continuous above a mass gap (for plus sign). 

In another development, in asymptotically flat spacetime, the boundary conditions associated with the KK soliton \cite{Witten:1981gj}, a double Wick rotation of the Schwarzschild black hole, makes the
KK vacuum with antiperiodic boundary conditions for the fermions
unstable towards decay, as the gravitational Hamiltonian is unbounded from below in this case. However, the AdS soliton 
\cite{Horowitz:1998ha}, which also has antiperiodic conditions for the fermions on an $S^1$, is perturbatively stable, though susy-
breaking. Recently it was shown that  supersymmetric AdS solitons exist \cite{Anabalon:2021tua,Anabalon:2022ksf, 
Anabalon:2024qhf, Anabalon:2022aig, Anabalon:2023oge}, and the charged solitons (for the AdS Einstein-Maxwell theory) generate 
phase transitions in the dual field theory. These ideas have also been generalized to 10 dimensions, representing new models of 
holographic confinement \cite{Canfora:2021nca, Nunez:2023nnl, Nunez:2023xgl,Fatemiabhari:2024aua}.

In this paper, we will find AdS soliton-like solutions in the well known STU model of type IIB supergravity. In general its field content is 
that of 3 $U(1)$ gauge fields and 2 scalars. It is a consistent truncation of the 5-dimensional maximal gauged supergravity that 
one gets from type IIB supergravity compactified on an $S^5$. As such, this solution 
should describe a deformation of ${\cal N}=4$ SYM, and we will find that there is a deformation of the Coulomb branch 
solution of \cite{Freedman:1999gk}, that interpolates between various possibilities for the spectrum, thus generating 
phase transitions in the field theory in 2+1 dimensions. For every possible value of the boundary sources, there are two possible AdS 
soliton like solutions \cite{Anabalon:2021tua}.  In the field theory, we find that there are two possible vacua in $\mathcal{N}=4$ Super 
Yang-Mills when the fermions are anti-periodic on an $S^1$. Thus, the solitons nicely describe this degeneracy and holography yields 
the strongly coupled phase diagram of $\mathcal{N}=4$ Super Yang-Mills in the large $N$ limit.

The paper is organized as follows. In section 2 we describe the model and the supersymmetric solutions. In section 3 
we describe the general solutions, have a first go at a field theory interpretation, and the parametrization of the space 
of solutions. In section 4 we describe holographic renormalization and describe the phase diagram from the point of view 
of gravity. Then we show that when the fermions are anti-periodic on the $S^1$, one can have two possible states in the dual  
$\mathcal{N}=4$ Super Yang-Mills, and we match these two results. In section 5 we uplift the solution to 10 dimensions, describe the 
result in terms of deformations of 
distributions of D3-branes, and analyze the mass spectra in order to obtain a field theory 
interpretation of the phase transitions. In section 6 we conclude, and the appendices give details on 
solving a relevant equation and the integrability conditions for the supersymmetry transformations. We also have an appendix on a 
possible interpretation of the solutions in terms of the Wick rotation of rotating D3-branes in 10 dimensions.

\section{The model}

We are interested in studying a truncation of type IIB supergravity compactified over the $S^5$ with action 
\begin{equation}
S_0=\frac{1}{2\kappa}\int\sqrt{-g}\left(  R-\frac{\left(
\partial\Phi_{1}\right)  ^{2}}{2}-\frac{\left(\partial\Phi_{2}\right)  ^{2}}{2}+\sum_{i=1}^{3}4L^{-2}X_i^{-1}
-\frac{1}{4}X_{i}^{-2}(F^{i})^{2}+\frac{1}{4}\epsilon^{\mu \nu \rho \sigma \lambda}A^1_{\mu}
F^2_{\nu \rho}F^3_{\sigma \lambda} \right)  d^{5}x,
\label{LSTU}%
\end{equation}
where $F^{i}$ are two forms, related with gauge fields in the standard
way, $F_{i}=d\bar{A}_{i}$, $X_{i}=e^{-\frac{1}{2}\vec{a}_{i}\cdot
\vec{\Phi}}$, $\vec{\Phi}=\left(  \Phi_{1},\Phi_{2}\right)  $ and%
\begin{equation}
\vec{a}_{1}=\left(  \frac{2}{\sqrt{6}},\sqrt{2}\right),\qquad\vec{a}_{2}
=\left(  \frac{2}{\sqrt{6}},-\sqrt{2}\right),\qquad\vec{a}_{3}=\left(  -\frac{4}{\sqrt{6}},0\right).
\end{equation}

We remark that we have changed the standard coupling constant of the gauged
supergravity by the AdS radius $L$ through the relation $g=\frac{1}{L}$. We
will be interested in purely magnetic solutions, in which case it is
consistent to truncate the axions to zero. The Lagrangian \eqref{LSTU} can be
obtained from the compactification of ten dimensional type IIB supergravity over the five sphere with the ansatz \cite{Cvetic:1999xp}
\begin{align}\label{uplift}
ds_{10}^{2}  &  =\tilde{\Delta}^{1/2}ds_{5}^{2}+L^{2}\tilde{\Delta}%
^{-1/2}\sum_{i=1}^{3}X_{i}^{-1}\left(  d\mu_{i}^{2}+\mu_{i}^{2}\left(
d\phi_{i}+\frac{1}{L}A_{i}\right)  ^{2}\right),\\
F_5&=G_5+*G_5,\\
G_5  &  =\frac{2}{L}\epsilon_{5}\sum_{i=1}^{3}\left(  X_{i}^{2}\mu_{i}%
^{2}-\tilde{\Delta}X_{i}\right)-\frac{L}{2}X_{i}^{-1}\ast_{5}dX_{i}\wedge d\mu_{i}
^{2}\\&+L^2\sum_{i}X_{i}^{-2}\mu_{i}d\mu_{i}\wedge\left(  d\phi_{i}+\frac
{1}{L} A_{i}\right)  \wedge\ast_{5}F_{i},%
\end{align}
where $*$ is the Hodge dual with respect to the ten-dimensional metric, 
$\ast_{5}$ is the Hodge dual with respect to the five-dimensional metric
$ds_{5}^{2}$, $\epsilon_{5}$ is its volume form, and $F_5$ is the self-dual five-form field
strength of type IIB supergravity. The $\phi_{i}$ are $2\pi$ periodic angular coordinates parametrizing the three 
independent rotations on $S^{5}$, $\tilde{\Delta}=\sum_{i}X^{i}\mu^2_i$ and  $\sum_{i}\mu^2_i=1$. We will be interested in
considering the higher-dimensional interpretation of some of our solutions
using this uplift.

The equations of ten-dimensional IIB supergravity in the metric-dilaton-$F_{5}$ sector are
given by%
\begin{eqnarray}
R_{\mu \nu }+2\nabla _{\mu }\nabla _{\nu }\phi_D -\frac{e^{2\phi_D }}{4}\left( 
\frac{1}{4!}F_{\mu \rho _{1}\dots \rho _{4}}F_{\nu }^{\ \rho _{1}\dots \rho
_{4}}-\frac{1}{2}g_{\mu \nu }\frac{1}{5!}F_{\rho _{1}\dots \rho _{5}}F^{\rho
_{1}\dots \rho _{5}}\right)  &=&0\ . \\
R-4\left( \partial \phi_D \right) ^{2}+4\square \phi_D  &=&0\ , \\
dF_{5} &=&0\ ,
\end{eqnarray}%
where $\phi_D$  is the ten dimensional dilaton. One then has to add by hand 
the self-duality condition $F_5=*F_5$. The lift of the solution has vanishing dilaton, and therefore the spacetime
is Ricci flat, which is consistent with the trace of the Einsteins' equations.

The Einstein's field equations in 5 dimensions are
\begin{align}
T_{\mu\nu}^{i}  &  =F_{\mu\rho}\,^{i}F_{\nu}^{i}{}^{\rho}-\tfrac{1}{4}%
g_{\mu\nu}F_{\rho\sigma}^{i}F^{i\rho\sigma}\,\,\,,\\
T_{\mu\nu}^{\Phi}  &  =\partial_{\mu}\Phi_{1}\partial_{\nu}\Phi_{1}+\partial_{\mu}\Phi_{2}\partial_{\nu}\Phi_{2}-g_{\mu \nu}\left(\frac{\left(
\partial\Phi_{1}\right)  ^{2}}{2}+\frac{\left(
\partial\Phi_{2}\right)  ^{2}}{2}-\sum_{i=1}^{3}4L^{-2}X_i^{-1} \right)  \,\,\,,\\
G_{\mu\nu} & =\frac{1}{2}T_{\mu\nu}^{\Phi}+\sum_{i=1}^{3}\frac{1}{2 X_{i}^2}T_{\mu\nu}^{i}\;,
\end{align}
plus the equations for the five-dimensional matter fields.

\subsection{Supersymmetry in type $D=5$ gauged supergravity}

The supersymmetry transformation of gravitino and the two dilatinos, that are the equations for the Killing spinor, for the $D=5$ gauged supergravity \eqref{LSTU} are \cite{Klemm:2000gh}

\begin{align}
\delta \psi _{\mu }dx^{\mu }&=\left( d+W\right) \Psi  =0 \ ,\\
\delta \lambda _{1}&=\sum_{i}\Omega _{i}\frac{\partial X_{i}}{\partial \Phi
_{1}}\Psi=0 \ ,\label{5D susy transformations} \\ 
\delta \lambda _{2}&=\sum_{i}\Omega _{i}\frac{\partial
X_{i}}{\partial \Phi _{2}}\Psi=0 \;,
\end{align}
where%
\begin{align}
A^{i} &= A_{\mu }^{i}dx^{\mu } \, \\
\Omega _{i} &=-\frac{1}{8}\left( X_{i}\right) ^{-2}\gamma
^{ab}F_{ab}^{i}-\frac{i}{4}\left( X_{i}\right) ^{-2}\left( \frac{\partial
X_{i}}{\partial \Phi _{1}}\slashed{\partial}\Phi _{1}+\frac{\partial X_{i}}{%
\partial \Phi _{2}}\slashed{\partial}\Phi _{2}\right) +\frac{i}{2L}\ , \\
W&=\frac{1}{4}\omega _{ab}\gamma ^{ab}-%
\frac{i}{2L}\sum_{i}A^{i}+\frac{i}{4!}\left( \gamma _{c}\gamma ^{ab}-6\delta
_{c}^{a}\gamma ^{b}\right) e^{c}\sum_{i}\left( X_{i}\right) ^{-1}F_{ab}^{i}+%
\frac{1}{3!L}\gamma _{c}e^{c}\sum_{i}X_{i}\ .  \notag
\end{align}%
The 1-form $e^c$ stands for the vielbein basis and $\omega_{a b}$ is the Levi-Civita spin connection 1-form. The complex spinor $\Psi $ is defined in terms of the symplectic
Majorana spinor $\epsilon ^{a}$ as $\Psi =\epsilon ^{1}+i\epsilon ^{2}$ (see
for instance \cite{Gutowski:2004yv}). We use the following basis for the Clifford
algebra:%
\begin{eqnarray}
\gamma ^{0} &=&-i\left( 
\begin{array}{cc}
0 & \sigma _{2} \\ 
\sigma _{2} & 0%
\end{array}%
\right) \ ,\quad \gamma ^{1}=-\left( 
\begin{array}{cc}
\sigma _{3} & 0 \\ 
0 & \sigma _{3}%
\end{array}%
\right) \ ,\quad \gamma ^{2}=i\left( 
\begin{array}{cc}
0 & -\sigma _{2} \\ 
\sigma _{2} & 0%
\end{array}%
\right) \ ,  \notag \\
\gamma ^{3} &=&\left( 
\begin{array}{cc}
\sigma _{1} & 0 \\ 
0 & \sigma _{1}%
\end{array}%
\right) \ ,\quad \gamma ^{4}=i\gamma ^{0}\gamma ^{1}\gamma ^{2}\gamma ^{3}\ .
\end{eqnarray}

The 2-form integrability conditions is defined as
\begin{equation}
    (dW+W\wedge W)\Psi =0.\label{5D integrability conditions}
\end{equation}

This equation leads a non-trivial solution only when the determinant of the components of $dW+W\wedge W$ is equal to zero.

\subsection{Supersymmetry in type IIB}
We are going to present new supersymmetric solutions in this theory. The general SUSY variations in the bosonic sector of type IIB is%
\begin{eqnarray}
\delta \lambda  &=&\frac{1}{2}\left( \Gamma ^{\mu }\partial _{\mu }\phi +%
\frac{1}{2}\slashed{H}_{3}\sigma _{3}\right) \epsilon -\frac{1}{2}e^{\phi_D
}\left( \slashed{F}_{1}i\sigma _{2}+\frac{1}{2}\slashed{F}_{3}\sigma _{1}\right)
\epsilon \ , \\
\delta \psi _{\mu }dx^{\mu } &=&d\epsilon +\frac{1}{4}\omega _{ab}\Gamma
^{ab}\epsilon +\frac{1}{4}\frac{1}{2!}H_{\mu ab}\Gamma ^{ab}dx^{\mu }\sigma
_{3}\epsilon  \\
&&+\frac{1}{8}e^{\phi_D }\left( \slashed{F}_{1}i\sigma _{2}+\slashed{F}_{3}\sigma _{1}+%
\frac{1}{2}\slashed{F}_{5}i\sigma _{2}\right) \Gamma _{\mu }dx^{\mu }\epsilon \;,
\notag
\end{eqnarray}%
where the slash for any $p-$form is defined as%
\begin{equation}
\slashed{F}_{p}=\frac{1}{p!}F_{a_{1}\dots a_{p}}\Gamma ^{a_{1}\dots a_{p}}\ .
\end{equation}

Note that in our configuration $\phi_D =0$ and $F_{1}=F_{3}=H_{3}=0$, then the
susy transformations are%
\begin{eqnarray}
\delta \psi _{\mu }dx^{\mu } &=&d\epsilon +\frac{1}{4}\omega _{ab}\Gamma
^{ab}\epsilon +\frac{1}{8}\frac{1}{2}\slashed{F}_{5}i\sigma _{2}\Gamma _{a
}e^{a }\epsilon \equiv D\epsilon \ .  \label{variation of 3/2 spinor}
\end{eqnarray}

The 2-form integrability conditions obtained by computing the commutator of
the derivative defined in (\ref{variation of 3/2 spinor}), as it is explained in detail in Appendix B, are
\begin{equation}\label{integravilityconditionsIIB}
\Xi =\frac{1}{4}R_{ab}\Gamma ^{ab}+\frac{1}{16}\frac{1}{5!}i\sigma _{2}%
\mathcal{D}F_{b_{1}\dots b_{5}}\Gamma ^{b_{1}\dots b_{5}}\Gamma _{a}e^{a}-%
\frac{1}{128}\frac{1}{4!}\slashed{F}_{5}F_{ad_{1}\dots d_{4}}\Gamma ^{d_{1}\dots
d_{4}}e^{a}\wedge \Gamma _{c}e^{c}\ .  
\end{equation}

\section{New AdS soliton in Type IIB supergravity}

AdS soliton type solutions with magnetic fluxes where found in the minimal gauged supergravity in five dimensions in \cite{Anabalon:2021tua}. We shall generalize these solutions now by including a non-trivial scalar profile. These solutions are double analytic continuations of a particular case of the electrically charged black hole solutions of the $U(1)^3$ truncation of the maximal gauged supergravity in five dimensions theory \cite{Cvetic:1999xp}, which oxidize to spinning D3 branes in 10 dimensions.  The vierbein and matter fields are
\bea
e^{0}  &  =&\Omega(x)^{1/2}dt\text{ ,}\cr
e^{1}  &  =& \frac{\Omega(x)^{1/2}}{2 x [(x-1) F(x) \eta]^{1/2} }dx\text{ ,}\cr
e^{2}  &  =& \Omega(x)^{1/2}F(x)^{1/2} L d\phi\text{ ,}\cr
e^{3}  &  =& \Omega(x)^{1/2}dz\text{ ,}\cr
e^{4}  &  =& \Omega(x)^{1/2}dy\text{ ,}\cr
\Phi_1 &  =& \sqrt{\frac{2}{3}}\ln(x)\text{ ,}\cr
\Phi_2 &  =& 0\text{ ,}\cr
A^{1}  &  = &q_{1}\left(  x^{-1}-x_{0}^{-1}\right) L d\phi\text{ ,}\cr
A^{2}  &  = &q_{1}\left(  x^{-1}-x_{0}^{-1}\right)L  d\phi\text{ ,}\cr
A^{3}  &  = & q_{2}\left(  x-x_{0}\right) L  d\phi\text{ ,}
\eea
with
\bea\label{Fx}
\Omega(x)  &  =&\frac{x^{2/3}\eta}{x-1}\text{ ,}\cr
F(x)  &  =& 
L^{-2}+\frac{(-1+x)^2(q_{1}^2-q_{2}^2 x)}{\eta x^2}.
\eea

As we shall see, the conformal boundary of the metric is located at $x=1$. When the integration constant $\eta>0$, then the 
range of the coordinate $x$ is constrained to be $1\leq x \leq x_0$, with $F(x_0)=0$, the center of the spacetime 
(this would be a "horizon" if $F(x)$ would be in front of $-dt^2$). When $\eta<0$, then $x_0\leq x \leq 1$. 
These two cases are not diffeomorphic to each other, as the scalar field is either everywhere positive or negative depending 
on which case one considers. Therefore the above configuration describe two physically inequivalent physical situations. 

We should note that, in fact, there $F(x_0)=0$ does not always have solutions: 

-if $\eta<0$, then $q_2=0$ means there is an $x_0$, but $q_1=0$ means there isn't. Note, however, that even a 
very small $q_1$ is enough to guarantee that there is an $x_0$.

-if $\eta>0$, then $q_2=0$ means there is no $x_0$, but $q_1=0$ means there is one. Note, however, that even a 
very small $q_2$ is enough to guarantee that there is an $x_0$. 

-if $|q_1|=|q_2|=q$, so, as we shall see, this is the supersymmetric solution, 
then if $\eta<0$, there always is an $x_0$ (independently of $q$), but if $\eta>0$, for large $q$ there is a 
solution, but for small $q$ (and in particular for $q_1=q_2=0$) there isn't. 

The canonical form of an asymptotically locally AdS$_{5}$ spacetime is achieved with the transformation (valid for $\eta>0$, 
the other case corresponds to changing $\eta$ into $-\eta$)
\bea
x&=&1+\frac{\eta L^2}{\rho^2}+\frac{2 \eta^2 L^4}{3 \rho^4}+\frac{\eta^3 L^6}{3 \rho^6}+O(\rho^{-8}),\cr
\Omega(x)  &  =&\frac{\rho^{2}}{L^{2}}+O(\rho^{-4}),\cr
g_{\phi\phi}  &  =&\Omega(x)F(x)=\frac{\rho^{2}}{L^{2}}-\frac{\mu}{\rho
^2}+O(\rho^{-4}),\cr
g_{\rho\rho}  & =&\frac{L^{2}}{\rho^{2}}-\frac{\frac{2}{9}\eta^2 L^6-\mu L^4}{\rho
^6}+O(\rho^{-8}),\cr
\mu &  =&-\eta L^{4} (q_{1}^{2}-q_{2}^{2}).\label{mu}
\eea

\subsection{Supersymetric solution}

We shall prove now that the configuration with $q_{2}=-q_{1}$ is supersymmetric, using the five dimensional supersymmetric 
tranformations. However we show that once we uplift the configuration to type IIB SUGRA the configuration is also 
supersymmetric in the case $|q_1|=|q_2|$. One can see that when this is replaced in the integrability condition \eqref{5D integrability 
conditions}, its determinant is equal to zero. To integrate the
equations of the Killing spinor we introduce the radial coordinate $r$,
which is the same one as the one we will introduce later for the uplift to 10
dimensions, through the change of coordinate%
\begin{equation}
x=\left( 1+\epsilon \frac{\ell ^{2}}{r^{2}}\right) ^{-1}\ ,
\end{equation}%
where $\epsilon =\pm 1$, and $\ell$ is related to $\eta$ as $\eta =-\epsilon \ell ^{2}/L^{2}$. The five dimensional vielbeins 
that we will use are%
\begin{eqnarray}
e^{0} &=&\frac{r}{L}\lambda(r) dt\ ,
\qquad e^{1}=\frac{dr}{r \lambda(r)^{2}\sqrt{F(r)}}\ ,\quad e^{2}=\frac{r}{L}\lambda(r) dy \\
e^{3} &=&\frac{r}{L}\lambda (r) dz\ ,
\quad e^{4}=r\lambda(r) \sqrt{F(r) }d\phi \ , \\
F(r)  &=&\frac{1}{L^{2}}-\epsilon \frac{\ell ^{2}L^{2}}{r^{4}}\left( q_{1}^{2}-q_{2}^{2}\lambda \left( r\right) ^{-6}\right) \ ,
\qquad \lambda \left( r\right) ^{6}=1+\epsilon \frac{\ell ^{2}}{r^{2}}\ .
\end{eqnarray}%

From the supersymmetry transformations \eqref{5D susy transformations}, we integrate the
Killing spinors when $q_1=-q_2$, which gives two linearly independent complex spinors 
\begin{eqnarray}
\Psi _{1} &=&e^{-\frac{i\pi \phi }{\delta}+\sigma (r) }\left( 
\begin{array}{c}
1 \\ 
\epsilon \frac{\lambda \left( r\right) ^{3}r^{3}}{\ell ^{2}L^{2}q_{1}}\left(
LF\left( r\right) ^{1/2}-1\right)  \\ 
0 \\ 
0
\end{array}
\right) \ ,  \label{KS1} \\
\Psi _{2} &=&e^{-\frac{i\pi \phi }{\delta}+\sigma(r) }\left( 
\begin{array}{c}
0 \\ 
0 \\ 
1 \\ 
-\epsilon \frac{\lambda \left( r\right) ^{3}r^{3}}{\ell ^{2}L^{2}q_{1}}%
\left( LF\left( r\right) ^{1/2}-1\right) 
\end{array}%
\right) \ ,  \label{KS2}
\end{eqnarray}%
where%
\begin{equation}
\sigma (r) =\int_{1}^{r}\frac{\left( 1+2\lambda \left( u\right)
^{6}\right) \left( 3-2LF\left( u\right) ^{1/2}\right) }{6u\lambda \left(
u\right) ^{6}LF\left( u\right) ^{1/2}}du\ ,
\end{equation}%
and $\delta$ is the period of the coordinate $\phi $,
implying that the Killing spinor are anti-periodic. The presence of two
complex Killing spinors means that the solution is 1/8 BPS. As a
cross-check, we verify that in these conventions $AdS_{5}$ has four
independent complex Killing spinors, constructed as given in section 3.1 of \cite{Klemm:2000nj} within the 
$\mathcal{N}=2$ theory. The Killing spinors are anti-periodic in the coordinate $\phi $,
with period $\delta$. The most general Killing spinor is a linear
combination of (\ref{KS1}) and (\ref{KS2})%
\begin{equation}
\Psi =c_{1}\Psi _{1}+c_{2}\Psi _{2}\ ,  \label{most general KS}
\end{equation}%
with complex coefficients $c_{1}$ and $c_{2}$. The Killing vector constructed
from the Killing spinors (\ref{most general KS}) gives a combination of all the Killing vectors of the spacetime
\begin{eqnarray}
\Psi ^{\dagger }\gamma ^{0}\gamma ^{\mu }\Psi \partial _{\mu } &=&-L\left(
\left\vert c_{1}\right\vert ^{2}+\left\vert c_{2}\right\vert ^{2}\right)
\partial _{t}+\left( \left\vert c_{1}\right\vert ^{2}-\left\vert
c_{2}\right\vert ^{2}\right) \partial _{y}-\left( c_{1}^{\ast
}c_{2}+c_{1}c_{2}^{\ast }\right) \partial _{z}  \notag \\
&&+i\left( c_{1}^{\ast }c_{2}-c_{1}c_{2}^{\ast }\right) \partial _{\varphi
}\ .
\end{eqnarray}

\subsection{Dual interpretation: basic analysis}

Below we will make clear that $\mu$ is proportional to the energy of the configuration. The
expansion of the scalar field yields%
\begin{align}
\Phi_1 &  =\frac{\Phi_{0}}{\rho^2}+\frac{\sqrt{6} \Phi_{0}^2}{12 \rho^{4}
}+O(\rho^{-8}), \label{Phi1}\\
\Phi_{0}  &  =\frac{\sqrt{2} L^2 \eta}{\sqrt{3}}\text{ .}
\end{align}

Hence, these solitons excite a VEV of an operator of conformal dimension $\Delta=2$ in the dual field theory, 
more precisely, in terms of ${\cal N}=4$ SYM, 
the symmetric traceless operator in the $\underline{\bf 20}'$ representation of $SO(6)$, ${\rm Tr}[X^I X^J-\frac{1}{6}\delta^{IJ}X^2]$, 
restricted to the neutral singlet $(1,1)_0$ under the decomposition of $SO(6)\rightarrow SO(2)\times SO(4)\simeq 
SO(2)\times SO(3)\times SO(3)$. 

The case of operators with $\Delta=2$ in $d=4$ is very special and has to be treated separately, as considered in 
\cite{Bianchi:2001kw}, for the case of the "Coulomb Branch" (CB) flow of \cite{Freedman:1999gk} which, as we will 
shortly see, corresponds to our own solution (as our solution is a generalization of that flow).

In the standard case ($2\Delta-d\neq 0$, with $d$ the dimension of the spacetime where the conformal field theory is defined), the expansion of the scalar of mass $m=\sqrt{\Delta(\Delta-d)}/R$ 
in terms of $z=R^2/\rho$ is \cite{Skenderis:2002wp}
\be
\Phi=z^{d-\Delta}\left[\phi_{(0)}+z^2\phi_{(2)}+...+z^{2\Delta-d}\left(\phi_{(2\Delta-d)}+\log z^2 \tilde\phi_{(2\Delta-d)}\right)+...\right]\;,
\ee
where the independent coefficients are: the non-normalizable mode $\phi_{(0)}$, corresponding to the operator source in the 
dual, and $\phi_{(2\Delta-d)}$, sometimes also called $\phi_{(1)}$, corresponding to the operator VEV. $\phi_{(2)},...,
\tilde \phi_{(2\Delta-d)},...$ are dependent on $\phi_{(0)}$, for instance 
\bea
\tilde \phi_{(2\Delta-d)}&=& -\frac{1}{2^{2\Delta-d}\Gamma\left(\Delta-\frac{d}{2}\right)\left(\Delta-\frac{d-2}{2}\right)}(\d_i\d_i)^{\Delta
-\frac{d}{2}}\phi_{(0)}\cr
\phi_{(2)}&=& \frac{1}{2\left(2\Delta-d-2\right)}\d_i\d_i\phi_{(0)}\;,
\eea
while $\phi_{(2\Delta-d)}$ gives the operator VEV by 
\be
\langle {\cal O}\rangle_{\phi_{(0)}}=-(2\Delta-d)\phi_{(2\Delta-d)}+F(\phi_{(0)})\;,
\ee
with $F$ a scheme-dependent function.

But when $\Delta=d/2$ like in our case ($\Delta=2, d=4$), one has to treat things separately, since there are several 
zero prefactors in the above, and as we see, $\phi_{(0)}$ and $\phi_{(2\Delta-d)}$ (normally the source and VEV) 
appear at the same order in the expansion. Another way to see this is that the mass formula has a double root, at the 
saturation of the BF bound ($m^2R^2\geq -d^2/4$): $m^2R^2=(\Delta-d/2)^2-d^2/4$. The expansion in our case 
($d=4,\Delta=2$) is, instead,
\be
\Phi=z^2\left[\log z^2\left(\phi_{(0)}+z^2 \phi_{(2)}+z^2\log z^2\psi_{(2)}+...\right)+\left(\tilde \phi_{(0)}
+z^2\tilde \phi_{(2)}+...\right)\right]\;,
\ee
where now $\phi_{(0)}$ is the operator source, and $\tilde\phi_{(0)}$ is the operator VEV. 

The expansion of the scalar 
in \cite{Bianchi:2001kw} coincides with our own (\ref{Phi1}) in the $\eta<0$ case. That means that there is no source, only an operator VEV $\Phi_0\propto
\eta$, 
parametrizing the Coulomb Branch (in \cite{Bianchi:2001kw}, the operator VEV was constant). In this $\Delta=d/2=2$ case, 
the prefactor $(2\Delta-d)$ of the operator is replaced by 2, so (since $\tilde\phi_{(0)}\equiv \Phi_0$ for us)
\be
\langle {\cal O}\rangle=2\tilde \phi_{(0)}\equiv 2\Phi_0=\frac{2\sqrt{2}L^2}{\sqrt{3}}\eta.
\ee

The solution we have found is a generalization of the Coulomb Branch case in \cite{Freedman:1999gk,Bianchi:2001kw}
both by the VEV parameter $\eta$ above, and by the parameters $q_1,q_2$ proportional to the boundary value of the gauge fields. Next we shall discuss the phase space of these solutions.

\subsection{The space of solutions}

A soliton solution is fully characterized in terms of its boundary conditions. Above we discused the solution in 
terms of the parameters $(\eta,q_1,q_2)$, the last two of which do not have a direct physical meaning ($\eta$ is the 
operator VEV in the field theory). A good set of physical variables are the boundary values of the gauge fields and the period 
$\phi\in [0,\delta]$. Indeed, solitons exist provided a regularity condition is imposed. This yields a boundary condition, namely 
the period $\delta$ of the $S^1$ is fixed by requiring the absence of conical singularities at $x_0$ (in the case when there is an 
$0<x_0$ such that $F(x_0)=0$, otherwise no soliton exists). 
In the Wick-rotated (in $t$) Euclidean black hole case, this would correspond to no singularities at the horizon, and would 
fix the temperature of the black hole. In the case at hand, one is actually working at zero temperature. Therefore, the scale 
is set by the KK scale $\delta$, for compactification of the 4-dimensional theory onto $\phi$, down to 2+1 dimensions. At this 
scale, the dimensionally reduced theory becomes just the 4-dimensional theory, KK expanded onto 2+1 dimensions.

The usual calculation, together with the condition $F(x_0)=0$, gives the period of the angle $\phi\in [0,\delta]$, 
with\footnote{Note that since $F(x_0)=0$ gives $\eta x_0^2=(x_0-1)^2(q_1^2-q_2^2x_0)$, we have $x_0=x_0(q_1,q_2,\eta)$.}  
\begin{equation}
\delta= \frac{ 2\pi x_0}{\lvert-q_2^2 x_0^2-q_2^2 x_0+2 q_1^2\rvert}\sqrt{	\left|\frac{-q_2^2 x_0+q_1^2}{-1+x_0}\right|}
=\frac{2\pi x_0^2\sqrt{\eta/L^2}}{\lvert -q_2^2x_0^2-q_2x_0+2q_1^2\rvert \lvert x_0-1\rvert^{3/2}}\;.
 \label{delta}
\end{equation}

Since  $x_0=x_0(q_1,q_2,\eta)$, it follows that, 
in the interpretation of the period of $\phi$ as inverse Kaluza-Klein temperature for compactification, we have, in terms of the 
previous set of parameters,
\be
\frac{1}{\delta}\equiv T_{\rm KK}=T_{\rm KK}(x_0,q_1,q_2,\eta)=T_{\rm KK}(q_1,q_2,\eta).
\ee

We want to understand how $T_{\rm KK}$ (governing the coupling of the KK reduced boundary 3 dimensional field theory) and 
$\eta$ (governing its operator VEV) vary, as the parameters of the bulk solution $(q_1,q_2,\eta)$ are varied.

It is difficult to calculate the general situation, so we restrict to the supersymmetric solution, with $|q_1|=|q_2|\equiv q$. Then 
\be
\frac{1}{T_{\rm KK}}=\delta=\frac{2\pi x_0}{q\lvert x_0^2+x_0-2\rvert}=\frac{2\pi}{\sqrt{\eta}}\frac{\sqrt{\lvert 1-x_0\rvert}}{x_0+2}\;,\;\;
q=\frac{x_0\sqrt{\eta}}{\lvert 1-x_0\rvert^{3/2}}\;,
\ee
which means that:

-$\delta\rightarrow \infty$, so $T_{\rm KK}\rightarrow 0$, $\Leftrightarrow$ $\eta\rightarrow 0$ at fixed $x_0$,
or $x_0\rightarrow \infty$ at fixed $\eta$, 
both of which imply $q\rightarrow 0$. 

-$\delta\rightarrow 0$, so $T_{\rm KK}\rightarrow \infty$, $\Leftrightarrow$ $\eta\rightarrow\infty$ at fixed $x_0$,
or $x_0\rightarrow 0$ at fixed $\eta$, 
both of which imply $q\rightarrow \infty$. 

Thus {\em the KK temperature $T_{\rm KK}$ is varied between 0 and $\infty$ by the variation of the $q$'s, allowed by the solution}. 

On the other hand, at fixed $q$, we have:

-$\eta\rightarrow 0$ gives $x_0\rightarrow 1$, so in turn $\delta\rightarrow \infty$, so $T_{\rm KK}\rightarrow 0$. 

-$\eta\rightarrow \infty$ gives $x_0\rightarrow 0$ (or $\infty$), so in turn $\delta \rightarrow 0$, so $T_{\rm KK}\rightarrow \infty$. 

Thus at fixed $q$, {\em $T_{\rm KK}$ tunes the VEV $\eta$, or the VEV $\eta$ 
tunes $T_{\rm KK}$}, which gives a {\em phase transition at 
$T_{\rm KK}=0$} between the (no VEV, no "horizon") and (VEV, "horizon") phases. 

We now note that, if we consider the 4-dimensional gauge coupling $g^2_{\rm YM}$ fixed, then $T_{\rm KK}$ can be 
exchanged for the 3-dimensional gauge coupling (the coupling of the dimensionally reduced theory), since
\be
g^2_{\rm 3d,YM}=g^2_{\rm YM}T_{\rm KK}.\label{TKK}
\ee

Then, instead of the interpretation of phase transition in KK temperature $T_{\rm KK}$, one has a phase transition in coupling, i.e., a 
{\em quantum critical phase transition}, happening at $g^2_{\rm 3d,YM}=0$.

We will reinforce this interpretation later, when describing the mass spectrum coming from the gravity dual.

Finally, we find that it is more convenient to parametrize the system in terms of the normalized 5-dimensional 
(gravity dual) gauge invariant "Wilson lines" 
$(\psi_1,\psi_2)$, integrated on a curve $C=\d \Sigma_2$\footnote{From the point of view 
of the 5-dimensional bulk; note that by using a 2-dimensional surface $\Sigma_2$ between infinity, $x=1$, 
and the origin, $x=x_0$, and Stokes' law, we could write this as $\int_{\Sigma_2}B_{yz}d\Sigma^{yz}\equiv \int_{\Sigma_2}\epsilon_{yzt\rho\phi}
\d_\rho A_\phi d\rho d\phi$, so would be some 5-dimensional generalization of magnetic flux, but would not correspond in 
the boundary 4 dimensions to magnetic flux, unlike the $AdS_4$ case.}, parametrized
by $\phi$ at the boundary $x=1$, defined as 
\begin{align}
\lim_{x \to 1} \oint A^{1}&= q_{1}\left(  1-x_{0}^{-1}\right) L \delta\equiv 2\pi L \psi_1\;, \cr
\lim_{x \to 1} \oint A^{3}&= q_{2}\left(  1-x_{0}\right) L \delta\equiv 2\pi L \psi_2\;.\label{Pol}
\end{align}

In terms of these sources it is very easy to see that the location of the supersymmetric solution discussed above, with $q_1=-q_2$, 
yields
\begin{equation}
x_0=\frac{\psi_2}{\psi_1}.
\end{equation}

Hence, we find that for every value of the pair $(\psi_1,\psi_2)$ there is one and only one supersymmetric soliton with $q_1=-q_2$.

More generally, we can use the definition of  $(\psi_1,\psi_2)$ to eliminate the integration constants $(q_1,q_2)$ in the definition 
of $\delta$, (\ref{delta}). This determines $x_0$ in terms of the sources $(\psi_1,\psi_2)$,
\begin{equation}
\psi_1^2x_0^3+(\psi_2^4+4\psi_1^4-4\psi_2^2\psi_1^2-\psi_2^2-\psi_1^2)x_0^2-\psi_2^2(4\psi_1^2-2\psi_2^2-1)x_0
+\psi_2^4=0.\label{delta2}
\end{equation}

We see that the advantage of this parametrization is that the dependence on $\delta$, which would be present in $F(x_0)=0$ in terms 
of the boundary values of the gauge fields, and was also present in the previous form $x_0=x_0(q_1,q_2,\eta)$, drops out. 
Hence, we have only a 2-parameter set $(\psi_1,\psi_2)$ defining $x_0$. Note that $F(x_0)=0$ in the previous form 
means $\eta=\eta(x_0,q_1,q_2)$,\footnote{Specifically, 
$\eta =(x_0-1)^2(q_1^2-q_2^2x_0)/x_0^2$.} but $x_0=x_0(\psi_1,\psi_2)$ from (\ref{delta2}), while $q_1=q_1(\psi_1,x_0,\delta)$ and 
$q_2=q_2(\psi_2,x_0,\delta)$ from their definition,\footnote{Specifically, $q_1=2\pi\psi_1/[(1-x_0^{-1})\delta]$ and 
$q_2=2\pi \psi_2/[(1-x_0^{-1})\delta]$.} which finally means that, up to some possible 
discrete choices, $\eta=\eta(\delta,\psi_1,\psi_2)$ . 
Indeed then, the general solution is completely characterized once we give 3 parameters $(\delta, \psi_1,\psi_2)$.
Note that {\bf in the $(\psi_1,\psi_2)$ parametrization}, {\em we can cover both the $\eta<0$ and the $\eta>0$ solutions}.

The cubic equation (\ref{delta2}) is solved in the Appendix A. We find that generically there are two solitons for each value of the 
pair $(\psi_1,\psi_2)$. In Fig.\ref{fig1} we plot possibles $x_{0i}$ (see Appendix A for their definition), 
as a function of $\psi_1$, with $\psi_2$ fixed. Note that for a fixed, say, $\psi_2$, there is a maximum $\psi_1$ for 
which there is a solution (that is consistent with the existence of a $0<x_0$, with $F(x_0)=0$).
 The different branches intersect at  infinity, $x=1$, 
 where they yield the soliton of the Einstein-Maxwell 
theory in five dimensions \cite{Anabalon:2021tua}.
The plot points towards the existence of a non-trivial phase diagram in the canonical ensemble, as $\psi_1$ and $\psi_2$ are 
varied. Indeed, on the gravity side, we can find the energy of the solution by $E=E(\delta,\psi_1,\psi_2)$, which will allow us to study the phase diagram of these solutions.

\begin{figure}[H]
\centering
\includegraphics[scale=0.5]{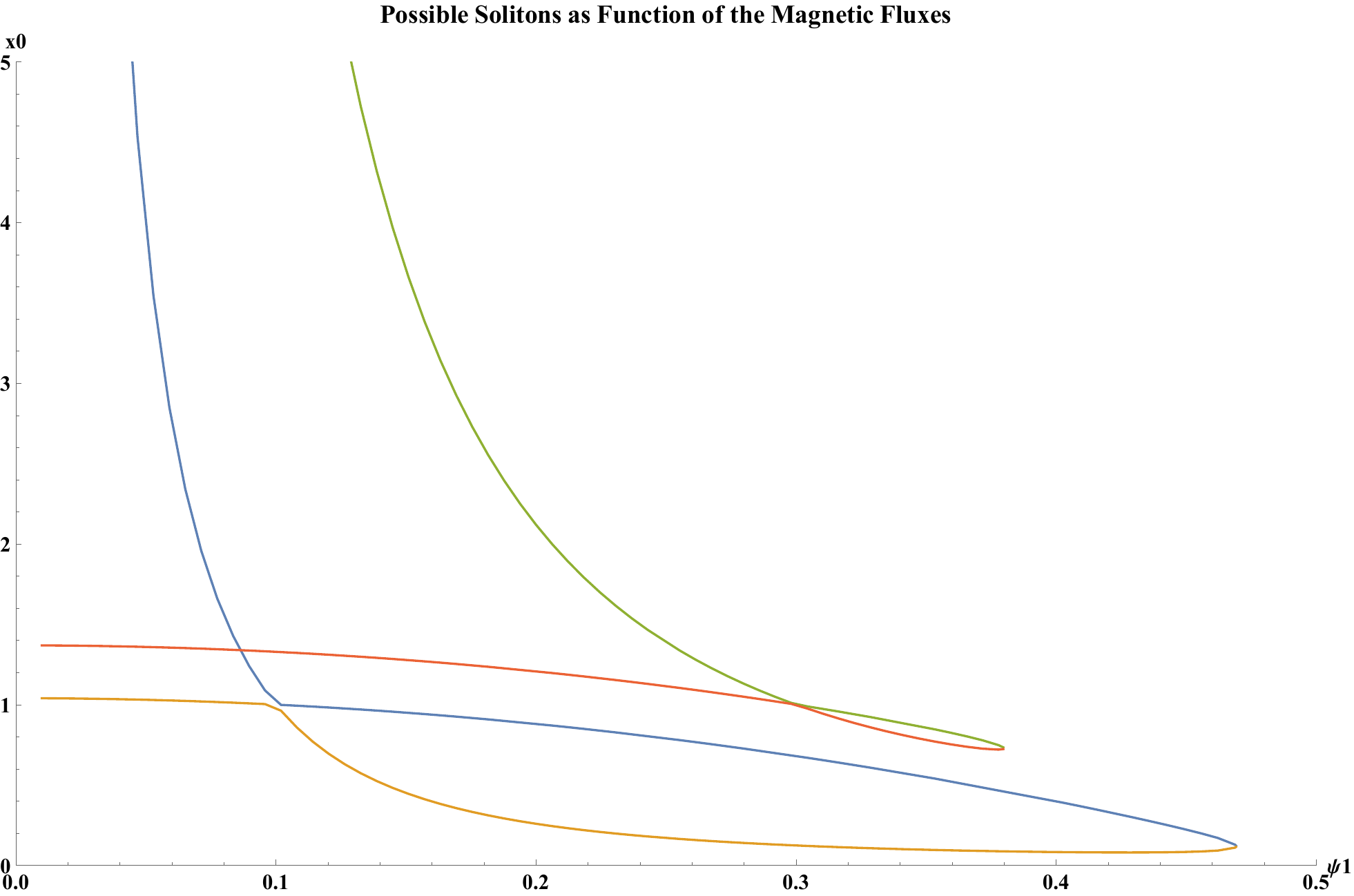}
\caption{The different colors are different physical roots of (\ref{Pol}). The $x_0$ in the $y$ axis are plotted vs the dimensionless 
Wilson line $\psi_1$ in the $x$-axis. The blue and yellow lines have $\psi_2=0.1$ and the red and green line have $\psi_2=0.3$. 
Either both solutions have a positive scalar field VEV or both have a negative scalar field VEV. The only roots that contribute to 
the physics
are $x_{01}$ and $x_{03}$ (see Appendix A for their definition).}
\label{fig1}
\end{figure}

As we saw in (\ref{TKK}), $T_{KK}=1/\delta$ defines the 3 dimensional gauge coupling, so
fixing $\delta$ is like fixing the coupling constant in the UV. Then, since we are working at zero temperature, {\em all }
the possible phase transitions are 
{\em quantum critical phase transitions}.

\section{Holographic renormalization and a phase diagram}

\subsection{Holographic renormalization}

Here we will use holographic renormalization to compute the expectation value of the dual energy momentum tensor. The countertems to deal with this situation were constructed in \cite{Balasubramanian:1999re, Bianchi:2001kw, Bianchi:2001de},
\begin{align}
 S=S_0+\frac{1}{\kappa}\int_{M^3\times S^1}K\sqrt{-h}d^4x+\frac{1}{2\kappa}\int_{M^3\times S^1}
 \sqrt{-h}\left(-\frac{6}{L}+\frac{1}{2 L}\left(\frac{1}{\ln (\rho/\rho_0)}-2\right)\Phi_1^2\right)d^4x   \;,
\end{align}
where $S_0$ is the action \eqref{LSTU} truncated to $\Phi_2=0$, $g_{\mu \nu}=h_{\mu \nu}+N_{\mu}N_{\nu}$, and 
$N_{\mu}$ is the outward pointing normal to the boundary and $K_{\mu \nu}=\frac{1}{2}\nabla_{\mu}N_{\nu}
+\frac{1}{2}\nabla_{\nu}N_{\mu}$ is the extrinsic curvature.  The boundary integrals are over the D3-brane geometry. 
Namely, a three dimensional Minkwoski spacetime times a circle,
\begin{equation}\label{bg}
ds^2=\gamma_{a b}dx^adx^b=-dt^2+dy^2+dz^2+d\phi^2    \;,
\end{equation}
which is the background spacetime for the quantum field theory. The scalar field has in general the asymptotic expansion 
\begin{equation}
\Phi_1=J_{\Phi}\frac{\ln(\rho^2/\rho_0^2)}{\rho^2}+\frac{\Phi_0}{\rho^2}+O\left(\frac{\ln(\rho^2/\rho^2_0)}{\rho^4}\right)\;,
\end{equation}
with the on-shell variation
\begin{equation}
\frac{\delta S}{\delta J_{\Phi}}=\frac{1}{2\kappa L^5}\Phi_0    .
\end{equation}

Indeed, our soliton has no scalar sources and this relation provides the holographic interpretation of 
$\Phi_0$ as a VEV, as already explained. The vacuum expectation value of the energy momentum 
tensor of the dual field theory is 
\begin{align}
\left<T_{a b}\right>&=\frac{-2}{\sqrt{-\gamma}}\frac{\delta S}{\delta \gamma^{a b}}\\
&=\lim_{\rho \to \infty}\frac{\rho^2}{L^2}\frac{-2}{\sqrt{-h}}\frac{\delta S}{\delta h^{a b}}\\
&=\lim_{\rho \to \infty}\frac{\rho^2}{L^2 \kappa}\left(h_{a b}\; K-K_{a b}-\frac{3}{L}h_{a b}-\frac{1}{2 L} h_{a b} \Phi_1^2 \right)\;,
\end{align}
which yields 
\begin{equation}
\left<T_{t t}\right>=-\frac{\mu}{2 L^3 \kappa}\;,\qquad\left<T_{z z}\right>=\left<T_{y y}\right>
=\frac{\mu}{2 L^3 \kappa}\;,\qquad \left<T_{\phi \phi}\right>=-\frac{3 \mu}{2 L^3 \kappa}\;.
\end{equation}

\subsection{Phase diagram from $E=E(\delta,\psi_1,\psi_2)$}

The free energy of the solitons in the canonical ensemble is just the energy. Hence we will be interested to see 
how the the energy changes when we vary the sources $(\psi_1,\psi_2)$. A convenient normalization of the energy 
is that of the AdS Soliton \cite{Horowitz:1998ha},

\begin{equation}
E_0=-\frac{L^3 \pi^4}{2\kappa\delta^3}V_2\;,
\end{equation}
where $V_2$ is the volume of the $y-z$ plane. So we plot the energy of the solution with the running scalar 
$E_\Phi=\left<T_{tt}\right>V_2 \delta$ divided by the absolute value of the energy of the AdS soliton in five dimensions,
\begin{align}
F_\Phi\equiv\frac{E_\Phi}{|E_0|}&=\left<T_{tt}\right>V_2 \delta \frac{2\kappa\delta^3}{L^3 \pi^4 V_2}\\
&=-16\frac{(\psi_1^2x_0^2-\psi_2^2)(\psi_1^2 x_0-\psi_2^2)}{x_0(x_0-1)^2}\;,
\end{align}

We note that for the supersymmetric solution with $q_1=-q_2$, we have
$\psi_1^2x_0^2-\psi_2^2=0=0$, so $F_\Phi=0$, as expected.
\begin{figure}[H]
\centering
\includegraphics[scale=0.4]{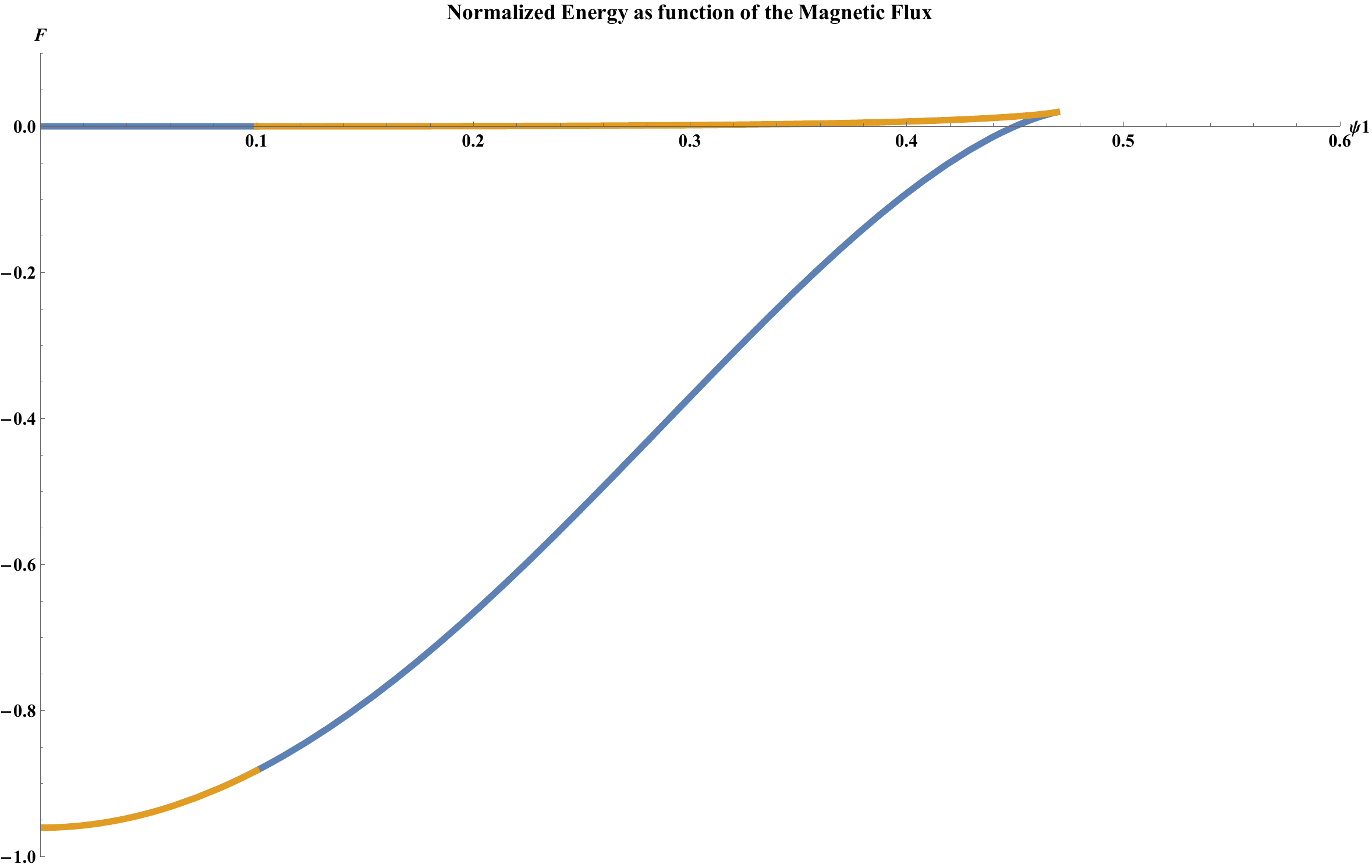}
\caption{The normalized energy as a function of the Wilson line $\psi_1$ when $\psi_2=0.1$. The phase diagram is 
composed by the different D3-brane distributions. They turn out to be continuously connected on the gauge theory 
side due to the introduction of the Wilson lines in 5 dimensions. The different roots of the polynomial (\ref{delta2}) 
have different colours.}
\label{fig2}
\end{figure}

\begin{figure}[H]
\centering
\includegraphics[scale=0.4]{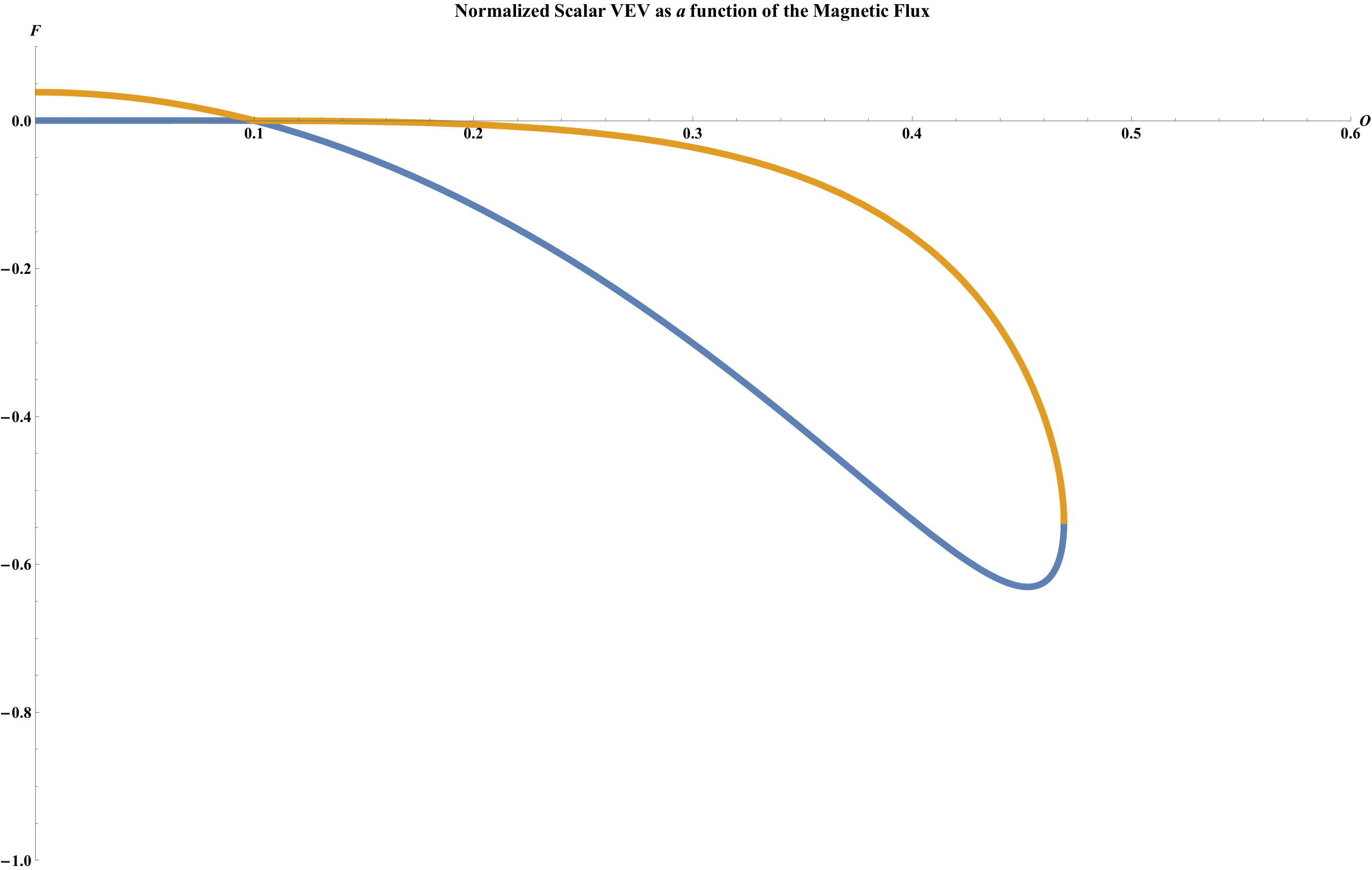}
\caption{The normalized scalar field vacum expectation value as a function of the Wilson line $\psi_1$, when $\psi_2=0.1$. 
Here we see that the VEV is negative for some solutions and positive for others. The negative VEV yields D3 brane 
distributions different than the positive VEV, as we will discuss below. There is a crossover between the different regimes.}
\label{fig3}
\end{figure}

The free energy $F_\Phi$ changes its color in Fig.\ref{fig2} in a continuous way. At this point is possible to see that the scalar 
VEV continuously goes to zero indicating a redistribution of the D3-branes in the sense of \cite{Freedman:1999gk}. 
This happens when $\psi_2=\pm\psi_1$. It is possible to see that the energy and all its derivatives are continuous at this point.
That means that there is continuous phase transition at $\psi_2=\pm \psi_1$, 
generically followed by the phase transition at ($q_1=\pm q_2$, so) $|\psi_2|=x_0 |\psi_1|<|\psi_1|$.

From $F(x_0)=0$, meaning $\eta x_0^2=(x_0-1)^2(q_1^2-q_2^2x_0)$, it is clear that we have the scalar VEV 
$\eta=0$ only for $q_1^2=q_2^2x_0$, 
meaning for $\psi_1^2x_0=\psi_2^2$, or for $x_0=1$, or  for both, in which case we have $\psi_1=\pm \psi_2$ and $x_0=1$, 
where the horizon disappears (so there we transition between the "horizon" and "no horizon" phases, as we already explained). 
This is {\em the "quantum phase transition" at $T_{KK}=0$ or $g^2_{\rm 3d,YM}=0$} described before. 

The solution on the lower branch (with $F_\Phi<0$) 
increases $\psi_i$ until at some $\psi_i$, one reaches $F_\Phi=0$, 
corresponding to the supersymmetric solution ($q_1=-q_2$). 
There we have a phase transition to the phase dominated by the D3 brane distributions of \cite{Freedman:1999gk} 
(which has zero energy), with anti-periodic boundary conditions for the fermions in $\phi$. 
From the point of view of the dual field theory, reduced on $\phi$ to 
3 dimensions, {\em this is another "quantum phase transition", at nonzero $g^2_{\rm 3d, YM}$}. 
One should note however that the distributions of \cite{Freedman:1999gk} are singular in the IR, 
so its inclusion in the phase diagram suppose that they actually become regular when quantum corrections are included.

\subsection{QFT energy}

Here we will discuss in greater detail how to understand the phase diagram from the QFT point of view. It is straightforward to compute the vacuum expectation value of the energy of a single scalar field in the background (\ref{bg}). The result is 
\begin{equation}
\langle E_{\rm QFT}\rangle=-\frac{\pi^2}{6\delta^3 } V_2 X   \;,
\end{equation}
where $X$ is a numerical factor that depends whether the scalar field is periodic or anti-periodic in the $S^1$. It comes from Riemann zeta-function regularization of the sum over the modes in the circle and it yields

\begin{align}
X_{even}=\sum_{n=1}^{\infty} (2n)^{3}=\frac{8}{120}\, ,\\   
X_{odd}=\sum_{n=1}^{\infty} (2n-1)^{3}=-\frac{7}{120}\, .
\end{align}

The field content of $\mathcal{N}=4$, $SU(N)$ super Yang-Mills is 6 scalars and 4 Weyl fermions in the adjoint representation 
plus one gauge vector. For the fermions the signs of the periodic and antiperiodic energies get interchanged. At weak coupling, 
we get the total energy by multiplying the scalar field energy by the number of degrees of freedom associated to each field,
 with the corresponding numerical factor depending on whether the fields are periodic or anti-periodic on the $S^1$. 
 So for the case where the scalars, the vectors and the fermions are antiperiodic, we get
\begin{equation}
\langle E_{\rm SYM}\rangle=-\frac{\pi^2\, V_2}{6\delta^3}X_{odd}(N^2-1)(6+2-8)=0.
\end{equation}

Hence this energy is automatically zero on account of the matching of the bosonic and fermionic degrees of freedom, and the fact that all fields have the same boundary condition on the $S^1$.

When the fermions are anti-periodic but the scalars and the vectors are periodic we get 
\begin{equation}
\langle E_{\rm SYM*}\rangle=-\frac{\pi^2\, V_2}{6\delta^3}X_{even}(N^2-1)(6+2+8\frac{7}{8})=-\frac{\pi^2\, V_2}{6\delta^3}(N^2-1).
\end{equation}

The AdS/CFT dictionary tell us that $\frac{L^3}{\kappa}=\frac{N^2}{4\pi^2}$. So the gravitational energy is
\begin{equation}
E_0=-\frac{\pi^2\, V_2}{8\delta^3}N^2=\frac{3}{4}\langle E_{\rm SYM*}\rangle\;,
\end{equation}
which is a well-known result valid at large $N$. Thus, we learn that in our phase diagram is possible to see the interplay of 
$\langle E_{\rm SYM*}\rangle$ and $\langle E_{\rm SYM}\rangle$, and that morevoer, the value of $\langle E_{\rm SYM}\rangle$ 
at strong coupling and vanishing sources is also zero, see Fig. \ref{fig2}. This explains from the field theory point of view the existence 
of two branches of gravity solutions.

\section{Continuous distributions of D3-branes vs. rotating D3-branes}

We start by reviewing some of the findings of \cite{Freedman:1999gk} on distributions of D3-branes. 
We show that when the gauge fields vanish in our soliton solutions, we recover the two different distributions 
of D3-branes that break the isometries of the $S^5$ to $SO(4)\times SO(2)$. The distribution of D3-branes 
of \cite{Freedman:1999gk} are solutions of the supergravity action

\be
I=\frac{1}{2\kappa}\int\sqrt{-g}\left(R-2\sum_{i=1}^{5}\left(
\partial\alpha_i\right)  ^{2}-V\right)  d^{5}x\;, \label{L2}
\ee
with
\bea
V&=&-\frac{1}{2 L^2}\left[{\rm Tr}(M)^2-2{\rm Tr}(M^2)\right],\cr
M&=&diag(e^{2\beta_1},e^{2\beta_2},e^{2\beta_3},e^{2\beta_4},e^{2\beta_5},e^{2\beta_6}),\cr
\vec{\beta}&=&\frac{1}{\sqrt{2}} B\vec{\alpha},
\eea
and
\begin{equation}
B=\begin{pmatrix}
1 & 1 & 1& 0&3^{-1/2}\\
1 & -1 & -1& 0&3^{-1/2}\\
-1 & -1 & 1& 0&3^{-1/2}\\
-1 & 1 & -1& 0&3^{-1/2}\\
0 & 0 & 0& \sqrt{2}&-\frac{2}{3^{1/2}}\\
0 & 0 & 0& -\sqrt{2}&-\frac{2}{3^{1/2}}\\
\end{pmatrix}.
\end{equation}

Here $M$ is a representative of the coset $SL(6,\mathbb{R})/SO(6)$, and the action of $SO(6)$ on $M$ is by conjugation. 
Note that $B^{T}B=4 \mathbb{1}_{5\times 5}$. Hence, the Lagrangian (\ref{L2}) is manifestly $SO(6)$ invariant. 

The case $n=2$ of table 1 of \cite{Freedman:1999gk} is recovered when $\vec{\alpha}=(0,0,0,0,-\frac{1}{2}\Phi)$ 
in terms of a canonically normalized scalar field $\Phi$, and then $\vec{\beta}=-\frac{\Phi}{2\sqrt{6}}(1,1,1,1,-2,-2)$. 
When the gauge fields vanish, this theory exactly coincides with the theory \ref{LSTU} when $\Phi=\Phi_1$. 
In the conventions of \cite{Freedman:1999gk}, this flow has $\Phi<0$, and therefore this corresponds in our 
coordinates to having $x<1$ and $\eta<0$.

The case $n=4$ of table 1 of \cite{Freedman:1999gk} corresponds to $\vec{\alpha}=(\frac{\sqrt{3}}{4}\Phi,0,0,0,\frac{1}{4}\Phi)$ 
with the canonically normalized scalar field $\Phi$, and then $\vec{\beta}=\frac{\Phi}{2\sqrt{6}}(2,2,-1,-1,-1,-1)$. 
In this case we match their potential with $\Phi=\Phi_1$. This flow has $\Phi>0$, which in our coordinates is $x>1$ and $\eta>0$.

\subsection{Uplift of the metric to 10 dimensions}

For the purposes of top-down AdS/CFT (whose rules are {\em derived} from string theory), 
it is not enough to consider a 5-dimensional solution; rather, one has to have a 10-dimensional solution, 
moreover obtained from a D-brane configuration. This is possible in our case.

Indeed, using the uplift (\ref{uplift}) we can write our solution, with non-vanishing gauge fields, as follows.

Considering the change of variable $x=\left( 1+\epsilon \ell
^{2}/r^{2}\right) ^{-1}$ and using the uplift (\ref{uplift}), we can write the 10-dimensional metric as
\begin{eqnarray}
ds_{10}^{2} &=&\frac{\zeta(r,\theta) r^{2}}{L^{2}}\left( \frac{L^{2}dr^{2}}{%
r^{4}F(r) \lambda(r)^{6}}+dx_{1,2}^{2}+F\left( r\right) L^{2}d\phi
^{2}\right)\\
&&+\frac{L^{2}}{\zeta(r,\theta)}\left\{ \zeta(r,\theta) ^{2}d\theta ^{2}+\lambda(r)
^{6}\sin ^{2}\theta \left( d\phi _{3}+L^{-1}A_{3}\right) ^{2}\right.   \cr
&&\left. +\cos ^{2}\theta \left[ d\psi ^{2}+\sin ^{2}\psi \left( d\phi
_{1}+L^{-1}A_{1}\right) ^{2}+\cos ^{2}\psi \left( d\phi
_{2}+L^{-1}A_{2}\right) ^{2}\right] \right\} \ , \\
F\left( r\right)  &=&L^{-2}+\frac{\ell ^{4}}{\eta r^{4}}\left(
q_{1}^{2}-q_{2}^{2}\lambda(r)^{-6}\right) \ ,\qquad \lambda(r) ^{6}=1+\epsilon 
\frac{\ell ^{2}}{r^{2}}\ ,\qquad \zeta(r,\theta) ^{2}=1+\epsilon \frac{\ell ^{2}}{r^{2}%
}\cos ^{2}\theta\;,\cr
A_1&=&A_2=\epsilon q_1\ell^2\frac{r^2-r_0^2}{r^2r_0^2}Ld\phi\;,\qquad
A_3=\epsilon q_2\ell^2\frac{r^2-r_0^2}{(r^2+\epsilon\ell^2)(r_0^2+\epsilon \ell^2)}Ld\phi \;,\label{10dsol}
\end{eqnarray}%
where $\epsilon =\pm 1$ depending whether the scalar is positive or
negative, and $r_0$ is the zero of $F(r)$.  For consistency, $\eta $ and $\epsilon $ must have
opposite signs, hence we considered $\eta =-\epsilon \ell ^{2}/L^{2}$. 
We use $\vec{\mu}=\left( \cos \theta \sin \psi ,\cos \theta \cos \psi ,\sin \theta\right)$. 
The field strength 5-form $F_{5}=G_{5}+\star G_{5}$ is defined in terms of $%
G_{5}$ given by
\begin{eqnarray}
G_{5} &=&\frac{2r^{3}\epsilon }{L^{4}\lambda ^{6}}\left[ \sin ^{2}\theta
+\lambda ^{12}\cos ^{2}\theta -\zeta ^{2}\left( 1+2\lambda ^{6}\right) %
\right] dr\wedge dt\wedge dy\wedge dz\wedge d\phi  \\
&&-\frac{\epsilon \lambda ^{\prime }r^{3}}{\lambda L^{2}}\left[
2r^{2}\lambda ^{6}F\left( r\right) +3L^{2}\left( 1-\lambda ^{6}\lambda
_{0}^{-6}\right) \left( q_{2}^{2}-q_{1}^{2}\lambda ^{6}\lambda
_{0}^{6}\right) \right] \sin \left( 2\theta \right) d\theta \wedge dt\wedge
dy\wedge dz\wedge d\phi \notag \\ \notag
&&+3r^{3}\epsilon \lambda ^{5}\lambda ^{\prime }\left[ \sin \left( 2\theta
\right) d\theta \wedge \left( q_{1}\sin ^{2}\psi d\phi _{1}+q_{1}\cos
^{2}\psi d\phi _{2}+q_{2}d\phi _{3}\right) \right.  \\
&&\left. -q_{1}\cos ^{2}\theta \sin \left( 2\psi \right) d\psi \wedge \left(
d\phi _{1}-d\phi _{2}\right) \right] \wedge dt\wedge dy\wedge dz\ .\notag
\end{eqnarray}

The field strength 5-form can be written explicitly as the exterior derivative of a 4-form as 
$F_5=d(\mathcal{C}_4+\tilde{\mathcal{C}}_4)$, where 
\begin{eqnarray}
\mathcal{C}_{4} &=& -\left[\frac{r^{4}}{L^{4}}\zeta(r,\theta)^{2}+\frac{\ell^{4}}{r_{0}^{2}}\epsilon\cos^{2}\theta(q_{2}^{2}\lambda(r_{0})^{-6}-q_{1})\right]dt\wedge dy\wedge dz\wedge d\phi\\
 & &+\ell^{2}\epsilon(q_{1}\cos^{2}\theta(\cos^{2}\psi d\phi_{2}+\sin^{2}\psi d\phi_{1})+q_{2}\cos^{2}\theta d\phi_{3})\wedge dt\wedge dy\wedge dz\,\notag\\
\tilde{\mathcal{C}}_{4} &=&\frac{L^{4}r^{2}\lambda(r)^{6}\cos^{4}\theta\sin(2\psi)}{2r^{2}\zeta(r,\theta)^{2}}d\phi_{1}\wedge d\phi_{2}\wedge d\phi_{3}\wedge d\psi\\
 &&-\frac{L^{4}q_{2}r^{2}(\lambda(r)^{6}-\lambda(r_{0})^{6})}{r^{2}\zeta(r,\theta)^{2}\lambda(r_{0})^{6}}\cos^{4}\theta\cos\psi\sin\psi d\phi\wedge d\phi_{1}\wedge d\phi_{2}\wedge d\psi \notag\\
 && -\frac{\ell^{2}L^{4}q_{1}r\epsilon\cos(2\psi)\sin^{2}(2\theta)}{8r^{4}\zeta(r,\theta)^{4}}(1+\frac{\epsilon\ell^{2}}{r_{0}^{2}}\cos^{2}\theta)dr\wedge d\phi\wedge\phi_{3}\wedge(d\phi_{1}-d\phi_{2}) \notag\\
 && -\frac{L^{4}r^{4}q_{1}\ell^{2}\epsilon\sin(2\theta)}{4r^{6}\zeta(r,\theta)^{4}}d\theta\wedge d\phi\wedge d\phi_{3}\wedge \notag\\
 && \left[-\zeta(r,\theta)^{4}(d\phi_{1}+d\phi_{2})+r^{2}\lambda(r)^{6}(r^{-2}-r_{0}^{-2})\cos(2\psi)\cos^{2}\theta(d\phi_{1}-d\phi_{2})\right]\notag
\end{eqnarray}%
with $G_5=d \mathcal{C}_4$, $*G_5=d\tilde{\mathcal{C}}_4$.

The flux of the $F_5$ on the $S^5$ with coordinates $[\theta,\psi,\phi_1,\phi_2,\phi_3]$ and ranges $\theta,\psi \in [0,\pi/2]$, $\phi_1,\phi_2,\phi_3\in [0,2 \pi]$  is given by
\begin{equation}
    \int_{S^5}F_5= \int_{S^5}\star F_5=\epsilon 4\pi^3 L^4 \, .
\end{equation}

Regarding the supersymmetry of the configuration, we show that the determinant of the components of integrability conditions 
\eqref{integravilityconditionsIIB}  are all zero for $|q_1|=|q_2|$, which ensures the existence of a solution of the Killing spinor equation 
\eqref{variation of 3/2 spinor}. Consequently, from the point of view of IIB supergravity, the Killing spinor equation admits a solution 
even in the case $q_1=q_2$, in addition to the case $q_1=-q_2$ that we found in $D=5$.

As we already mentioned, when the supergravity $U(1)$ gauge fields vanish, we recover the singular distributions of \cite{Freedman:1999gk}. These singularities are considered to be ``good'' in the analysis of \cite{Gubser:2000nd}. As remarked in \cite{Gubser:2000nd} these Coulomb branch states do not seem to admit a finite temperature analogue (without $U(1)$ gauge fields). However, the singularities satisfy the more general Gubser-criterion that the evaluation of the scalar field potential on the solution should never yield $ +\infty$. Indeed, this is a property of the STU-model of maximal supergravity which has a scalar field potential which is everywhere negative.

\subsection{Mass spectrum and phase transitions} 

In the case $A^i=0$ of \cite{Freedman:1999gk}, it was noted that for the dilaton,
one can reduce the 10-dimensional equation of motion onto the 5-dimensional one, if we have a warped product form,
\be
ds^2_{10}=\Delta^{-2/3}(r,\mu_i,\phi_i)ds^2_5(y,z,t,r,\phi)+ds^2_K(\mu_i,\phi_i,r)\;,\label{warped}
\ee
so $ds_K$ that can depend on $ds_5$, but $ds_5 $ independent on $ds_K$, and 
{\em if the dilaton is independent on $K$, so $\Phi=\Phi(t,y,z,r,\phi)$.} Here $\Delta =\sqrt{\det g_K/\det g_K^{(0)}}$, where 
$g_K^{(0)}$ is the metric of the {\em undeformed by $ds_5$} metric of $K$, i.e., in this case, the metric of the round $S^5$ sphere, 
and $g_K$ is the full deformed metric on $K$.

That is so, since we can easily verify that the 10-dimensional d'Alembertian operator on $\Phi$ is 
\begin{equation}
\Box_{10D}\Phi=\frac{\Delta^{2/3}}{\sqrt{-g}}{\partial_{\mu}(g^{\mu \nu}\sqrt{-g}\partial_{\nu}\Phi)}+ds_K\; terms\, ,
\end{equation}
where $g_{\mu \nu}$ is the 5-dimensional metric for $ds_5$. Hence, this is equivalent to solving the  d'Alembertian (massless
KG) equation in the 5-dimensional metric $ds_5^2$.

In our case, with $A^i\neq 0$, specifically $g_{\phi K}\neq 0$, we have the same situation, if we impose the additional constraint
that $\Phi$ is independent on the circle (KK) coordinate $\phi$, so $\Phi=\Phi(t,y,z,r)$ only, in which case we have the same 
5-dimensional $\Box $ operator, but acting on a field that only depends on 4 dimensions, so on the zero mode for the KK expansion 
on $S^1$.

By comparing this form with our own uplift form (\ref{uplift}), we see that 
\be
\tilde \Delta^{1/2}=\Delta^{-2/3}=\frac{\zeta(r,\theta)}{\lambda^2(r)}\;,
\ee
which means that the 5-dimensional metric in our case can be put into the form
\bea
ds^2_5&=&\frac{\lambda^2 r^2}{L^2}\left(\frac{L^2dr^2}{r^4F(r)\lambda^6}+d\vec{x}^2_{1,2}+F(r)L^2d\phi^2\right)\cr
&=& \frac{r^2}{L^2}\left(1+\epsilon\frac{\ell^2}{r^2}\right)^{1/3}\left[\frac{L^dr^2}{r^4F(r)\left(1+\epsilon\frac{\ell^2}{r^2}\right)}
+d\vec{x}^2_{1,2}+F(r)L^2d\phi^2\right]\;,
\eea
and by using redefining $r/L=L/z$, we have 
\be
ds_5^2=\frac{L^2}{z^2}\left(1+\epsilon\frac{\ell^2 z^2}{L^4}\right)^{1/3}\left[\frac{dz^2}{L^4F(z)\left(1+\epsilon
\frac{\ell^2z^2}{L^4}\right)}+d\vec{x}_{1,2}^2+F(z)L^2d\phi^2\right]\;,
\ee
with 
\be
F(z)=1-\epsilon\frac{\ell^2z^4}{L^6}\left(q_1^2-\frac{q_2^2}{1+\epsilon\frac{\ell^2z^2}{L^4}}\right).
\ee

Then the spectrum of the scalar $0^{++}$ glueballs is given by the eigenstates of the d'Alembertian operator in this 
5-dimensional background. 
Since 
\be
\Box \Phi=\frac{z^5}{\left(1+\epsilon\frac{\ell^2 z^2}{L^4}\right)^{1/3}}\d_z\left[\frac{\left(1+\epsilon\frac{\ell^2 z^2}{L^4}\right)}{
z^3}F(z)\d_z\right]\Phi+\frac{z^2}{\left(1+\epsilon\frac{\ell^2 z^2}{L^4}\right)^{1/3}}\d_i\d_i\Phi\;,
\ee
under the redefinition of the variable,
$dz=\sqrt{F(z)\left(1+\epsilon\frac{\ell^2 z^2}{L^4}\right)} du$, and of the function, 
with a $e^{i\vec{k}\cdot\vec{x}}$ plane wave in the $y,z,t$ directions, and with $\vec{k}^2=-M^2$, 
\be
\Phi=e^{i\vec{k}\cdot\vec{x}}\frac{z^{3/2}}{F(z)\left(1+\epsilon\frac{\ell^2 z^2}{L^4}\right)}\Psi(z)\;,
\ee
from $\Box\Phi=0$ we get the one-dimensional Schr\"{o}dinger equation, 
\bea
&&-\frac{d^2\Psi(u)}{du^2}+V(u)= M^2\Psi(u)\cr
&&V(z)= -\left[F(z)\left(1+\epsilon\frac{\ell^2 z^2}{L^4}\right)\right]^{1/4}z^{3/2}
\frac{d}{dz}\left\{\frac{F(z)\left(1+\epsilon\frac{\ell^2 z^2}{L^4}\right)}{z^3}\frac{d}{dz}\frac{z^{3/2}}
{\left[F(z)\left(1+\epsilon\frac{\ell^2 z^2}{L^4}\right)\right]^{1/4}}\right\}.\cr
&&
\eea

We see that we can redefine $\tilde F(z)\equiv F(z)\left(1+\epsilon\frac{\ell^2 z^2}{L^4}\right)$, in which case we are 
back to the case considered in \cite{Anabalon:2023lnk}.

We can now make the same analysis from before (in 5 dimensions, in terms of the $x$ coordinate) for the $z_0$ solving 
$F(z_0)=0$, but now also consider together with the one solving $\tilde F(z_0)=0$, which is more relevant:

-if $\epsilon=+1$, $q_2\rightarrow 0$ gives an $z_0$, but $q_1= 0$ gives no $z_0$ (but $q_1\rightarrow 0$, yet 
$q_1\neq 0$, gives an $z_0$). 

-if $\epsilon=-1$, $q_2=0$ gives no $z_0$ (but $q_2\rightarrow 0$, yet $q_2\neq 0$, gives an $z_0$), but $q_1\rightarrow 0$
gives an $z_0$. 

-if $q_1=\pm q_2$ (in the susy case), there always is an $z_0$.

-if $\epsilon=-1$, $q_2=0$, there is no solution to $F(z_0)=0$, {\em but there is a solution to $\tilde F(z)=0$, namely 
$z_0=L^2/\ell$}.

-if $\epsilon=+1$ and $q_1=q_2=0$, we get no $z_0$.

In the UV, at $z\rightarrow 0$, we have $\tilde F(z)\simeq 1$, so
we get $z\simeq u$ and the same potential for both $\epsilon=\pm1$, 
\be
V(u\simeq 0)\simeq \frac{15}{4u^2}\Rightarrow \Psi(u)=\sqrt{Mu}\left[C_1J_2(Mu)+C_2Y_2(Mu)\right].
\ee

In the IR, we can have a $z_0\neq 0$, or not, as we discussed, depending on $\epsilon,\ell$ and $q_1,q_2$. 

If we have a $z_0\neq 0$, then for $z\rightarrow z_0$, with $\tilde F(z)\simeq \tilde F'(z_0)(z-z_0)$, 
$u-u_{\rm max}\simeq 2\sqrt{(z-z_0)/\tilde F'(z_0)}$, and writing $u_{\rm max}=Kz_0$, we get 
\bea
V(u\simeq K z_0)&\simeq&-\frac{\tilde F'(z_0)}{16|z-z_0|} \simeq -\frac{1}{4(u-Kz_0)^2}\Rightarrow \cr
\Psi&\simeq& \sqrt{Kz_0-u}\left[C'_1J_0\left(M(u-Kz_0)\right)
+C'_2Y_0\left(M(u-Kz_0)\right)\right].
\eea

The IR boundary condition puts $C'_2=0$, so the $J_0$ solution continued to the UV (at $u=0$) must give also $C_2=0$, 
which will give a quantization condition on $M(Kz_0)=Mu_{\rm max}$, as $M=M_n$. But, of course, the quantization condition 
will depend on the parameters $(q_1,q_2,\ell)$ of the solution, which will define $u_{\rm max}$, decoupling the 
scale of $M_n$, $u_{\rm max}$, from the KK scale, $T_{\rm KK}=1/\delta$. In any case, the spectrum is discrete. 

On the other hand, if there is no $z_0$ (so $z_0=0$) in the IR, 

-if $\epsilon=+1, q_1=0$, 
then $\tilde F(z)\sim (q_2^2\ell^2/L^6)z^4$, so 
\be
V(z\rightarrow \infty)\simeq -\frac{1}{4}z^2\frac{q_2^2\ell^2}{L^6}\simeq -\frac{1}{4(u-u_{\rm max})^2}\;,
\ee
so, despite the fact that we don't have a $z_0$,  
we obtain the same form of the potential in terms of $u$, since $\int du \simeq 1/\sqrt{(q_2^2\ell^2/L^6)z^2}$. So again a discrete 
spectrum.

-if $\epsilon=-1, q_2=0$, then there is no $z_0$ for $F(z)$, but there is one for $\tilde F(z)$, so the solution is again the same 
as before, and we have a discrete spectrum.

-if $\epsilon=+1, q_1=q_2=0$, then there is no $z_0$ for $F(z)$ or $\tilde F(z)$, and then $\tilde F(z)\simeq \ell^2z^/L^4$, so
\be
V(z\rightarrow\infty)\simeq +\frac{\ell^2}{L^4}=V(u)\;,
\ee
so {\em we have a continuous spectrum above a mass gap at $M^2=\ell^2/L^4$}.

In conclusion, this case of $\epsilon=+1, q_1=q_2=0$ is the only one for which we have a qualitatively different spectrum. 

We can then say that 
{\em the introduction of the $q_1,q_2$ charges induces a phase transition from the spectrum continuous above a mass gap, 
continuously connected to the discrete spectrum}. At $q_1=q_2=0$, the two spectra seemed distinct, as they were
obtained in the two separate cases, $\epsilon=+1$ and $\epsilon=-1$, respectively. 

Finally, when we have the pure AdS space, obtained formally by putting $F(z)=1, \ell=0$, we obtain that the potential in 
the UV is valid everywhere, $u=z$ and $V(z)=\frac{15}{4 u^2}$. In this case, there is no limit on $u=z$, it spans from $u=0$ 
to $u=+\infty$, which means that the spectrum is continuous without a mass gap. 

Since, as we saw in section 4, we had two phase transitions, interpreted as quantum phase transitions from the point of view 
of the 3-dimensional dual field theory reduced on $\phi$, one from "no horizon" (given by the singular distributions of D3 branes)
 to "horizon" (at $g^2_{\rm 3d,YM}=0$), and then to "AdS space" (at $g^2_{\rm 3d, YM}\neq 0$), 
 these are: from continuous above a mass gap to discrete, to continuous without a mass gap.

\section{Discussion and conclusions}

In this paper we have found AdS solitons depending on three parameters, namely the two sources associated to the gauge 
fields, which were proportional to the charge parameters $q_1,q_2$, and the value of the periodicity of the circle $S^1$, $\delta$. 
We have shown that it is possible to describe the phase space in terms of the dimensionless sources $(\psi_1,\psi_2)$, together
with $\delta=1/T_{\rm KK}$. The solutions give a dual scalar VEV
$\langle {\cal O}\rangle_{(1,1)_0}$ in 3+1 dimensions, proportional to $\eta=\pm \ell^2/L^2$.
Among the solutions, a special role is played by the supersymmetric solutions, with $q_1=\pm q_2$.

We have found two phase transitions from the $(E,\psi_1,\psi_2)$ diagram, as $\psi_1$ is varied, one at $\psi_1=\pm \psi_2, 
x_0=1$, and another the one at $\psi_2=\pm \psi_1 x_0(\psi_1,\psi_2)$ and $E=0$, to the previous solutions of \cite{Freedman:1999gk}. 

Our set of solutions continuously connects all the possibilities described in \cite{Freedman:1999gk}. 
The 10-dimensional uplift of the solutions
was found to be a deformation of the D3-brane distributions of \cite{Freedman:1999gk}, and in the appendix below we 
hint towards its description as a system of D3-branes, obtained from the Wick rotation of the rotating D3-branes in 
3 independent planes, so one expects that there is a good string theory interpretation of the results, though we 
have not found it so far. 

In terms of the 2+1-dimensional interpretation, the supersymmetric solutions give a 
quantum critical phase transition, at $g^2_{\rm 3d, YM}=0$, between a phase with no VEV
(and no horizon in the dual), and spectrum that is continuous above a mass gap, 
and a phase with VEV (and horizon in the dual), and discrete spectrum, and the transition to periodic AdS space 
is to a continuous and no mass gap spectrum, at nonzero $g^2_{\rm 3d, YM}$. 

Remarkably enough, we have found that the phase diagram of these solutions should correspond to the strongly coupled 
description of the existence of two possible vacua of the large $N$ $\mathcal{N}=4$ SYM when compactified on an $S^1$ 
in four dimensions and antiperiodic boundary conditions for the fermions on the $S^1$. Unexpectedly, we found that at 
finite values of the source the supersymmetry breaking vacuum gets its supersymmetry restored, corresponding to the 
BPS states existing in supergravity.

Hence, this should correspond to the existence to a non-perturbative object in the field theory, most likely the Q-ball 
\cite{Coleman:1985ki}, embedded into the supersymmetric theory, and extended to strong coupling (where its stability properties
and mass value with respect to the ones fundamental fields are not currently understood). 
Indeed, we see that in the UV, at $x=1$, 
we have $A^1=A^2=q_1(1-x_0^{-1})Ld\phi$, $A^3=q_2(1-x_0)Ld\phi$.

In the case of the double Wick rotation of the solution, with $F(x)$ multiplying $-dt^2$ instead of $d\phi^2$ in the metric, 
this would give $A^1=A^2=q_1(1-x_0^{-1})Ldt$, $A^3=q_2(1-x_0)Ldt$, which is the standard case for $\mu_1=q_1(1-x_0^{-1})L$, $\mu_2=q_2(1-x_0)L$, 
chemical potentials, or sources, for the corresponding $U(1)$ charges $\int d^3x\; \rho$, where
$\rho\sim {\rm Tr}[\bar Z \d^0 Z]+{\rm Tr}[\bar\psi\gamma^0\psi]$, 
with $Z$ complex combinations of $X^I$'s, and $\psi$ complex fermions, both charged under the $U(1)$'s. 

Therefore in our case, $A^1=A^2=q_1(1-x_0^{-1})Ld\phi$ and $A^3=q_2(1-x_0)Ld\phi$, $\mu_1$ and $\mu_2$ are sources for the 
$U(1)$ current components in the $\phi$ direction, $\sim {\rm Tr}[\bar Z\d^\phi Z]+{\rm Tr}[\bar \psi\gamma^\phi\psi]$, so they are
understood as $J^\phi=\rho v^\phi=\rho\frac{d\phi}{dt}\equiv \rho\omega$ (if we had $\rho \vec{v}$, we would write 
$Z=Z(\vec{x}-\vec{v}t)$, with $\vec{x}=(y,z)$). We see that, {\em from the point of view of the reduced 2+1 dimensional 
theory in $(t,y,z)$, in which $\phi$ is an internal direction}, $\omega$ might be understood as Q-ball 
\cite{Coleman:1985ki} angular frequency (for effective potential $V_{\rm eff}(Z)=V+\frac{1}{2}\omega^2|Z|^2$), 
for writing $Z(\phi=\omega t,\vec{x})=e^{i\omega t}Z(\vec{x})$, and $J^\phi$ is then Q-ball charge density (except, of 
course, that we don't have a time dependence of the phase $\phi$, that was just assumed). Then $\mu_1,\mu_2$ would be
chemical potentials for the Q-ball charges. 

This might provide a generalization of the Coulomb Branch solution for ${\cal N}=4$ SYM, by the parameters $\eta$
(operator VEV) and $q_1,q_2$ (related to the $\mu_1,\mu_2$, the "chemical potentials for Q-ball charges"), 
that contains both solutions with arbitrary (or no) periodicity 
of $\phi$, better understood within ${\cal N}=4$ SYM, and solutions with periodic $\phi$ and cigar-type solution with an $x_0$
("horizon"), understood either from the point of view of the reduction to 3 dimensions ($(y,z,t)$), or from the point of view 
of Euclidean version of 4 dimensions, at finite KK temperature $T_{\rm KK}$. 
We expect to make this picture more concrete in a future work.

\section*{Acknowledgements}

The work of HN is supported in part by  CNPq grant 301491/2019-4 and FAPESP grant 2019/21281-4.
HN would also like to thank the ICTP-SAIFR for their support through FAPESP grant 2021/14335-0. The work of AA is supported in part by the FAPESP visiting researcher award 2022/11765-7 and the FONDECYT grants 1200986, 1210635, 1221504 and 1230853.

\newpage

\appendix

\section{ Solutions of the cubic equation \eqref{delta2}}

The solutions of \eqref{delta2} have the form $x_{0i}=\lambda_1 \cos\left(\Theta 
+ \frac{2\pi n_i}{3}\right)-\frac{\lambda_2}{3}$ with $n_i=0,1,2$ for $i=1,2,3$, and

\begin{align}
\lambda_1=&\frac{2}{3 \psi_1}\sqrt{8 \psi_1^4 \lambda_2-\psi_1^2 \lambda_2^2-8 \psi_2^2 \psi_1^2 \lambda_2
+2 \psi_2^4 \lambda_2+12 \psi_2^2 \psi_1^2-2 \psi_2^2 \lambda_2-2 \psi_1^2 \lambda_2-3 \psi_2^2-6 \psi_2^4}\;,\\
\lambda_2=&\frac{1}{\psi_1^2}\left(-4\psi_2^2 \psi_1^2+\psi_2^4+4 \psi_1^4-\psi_2^2-\psi_1^2\right)\;,\\
\Theta=&3^{-1}\arccos{\frac{4}{27\lambda_1^3 
\psi_1^2}\left(3\psi_2^2\lambda_2^2+3\psi_1^2\lambda_2^2+9\psi_2^2\lambda_2+18\psi_2^4\lambda_2
-12\psi_1^4\lambda_2^2-27\psi_2^4+12\psi_2^2\psi_1^2\lambda_2^2 \right.} \\
&\left.-3\psi_2^4\lambda_2^2-36\psi_2^2\psi_1^2\lambda_2+\psi_1^2\lambda_2^3\right).
\end{align}

\section{Integrability conditions}\label{Appendix. Integrability conditions general}

In this appendix we compute the integrability condition for IIB in the
metric-$F_{5}$ sector. The supersymmetry transformations of the spin 3/2
field is%
\begin{equation}
\delta \psi _{\mu }dx^{\mu }=d\epsilon +W\epsilon \equiv D\epsilon \ ,
\label{covD appendix}
\end{equation}%
where for our field content%
\begin{equation}
W=\frac{1}{4}\omega _{ab}\Gamma ^{ab}+\frac{1}{16}i\sigma _{2}\slashed%
{F}_{5}\Gamma _{a}e^{a}\ .  \label{def W appendix}
\end{equation}%

We define integrability conditions $2-$form as the commutator of the
covariant derivative defined in (\ref{covD appendix})%
\begin{equation}
\Xi \equiv D\wedge D\epsilon \ .
\end{equation}%

It is simple to show that%
\begin{equation}
\Xi =dW+W\wedge W\ .  \label{Xi in terms of W}
\end{equation}%

Let us compute it term by term. The exterior derivative of $W$ is%
\begin{equation}
dW=\frac{1}{4}d\omega _{ab}\Gamma ^{ab}+\frac{1}{16}i\sigma _{2}d\slashed%
{F}_{5}\Gamma _{a}e^{a}+\frac{1}{16}i\sigma _{2}\slashed{F}_{5}\Gamma
_{a}de^{a}\ .
\end{equation}%

Using the torsion-less condition $de^{a}+\omega _{\ b}^{a}\wedge e^{b}=0$
and the definition of curvature $2-$form $R_{\ b}^{a}=\omega _{\
b}^{a}+\omega _{\ c}^{a}\wedge \omega _{\ b}^{c}$ we obtain%
\begin{equation}
dW=\frac{1}{4}R_{ab}\Gamma ^{ab}-\frac{1}{4}\omega _{ac}\wedge \omega _{\
b}^{c}\Gamma ^{ab}+\frac{1}{16}i\sigma _{2}d\slashed{F}_{5}\Gamma _{a}e^{a}-%
\frac{1}{16}i\sigma _{2}\slashed{F}_{5}\Gamma _{a}\omega _{\ c}^{a}\wedge e^{c}\
.  \label{dW to replace}
\end{equation}%

Note that in general one can write $W=W_{A}\otimes \Gamma ^{A}$ where $W_{A}$
is the tensor product of the 1-form space and $2\times 2$ matrices, in
general we suppress the tensor product symbol. The repeated indices $A$ are
sumed over all terms which defines $W$ and encodes the index structure of
the $\Gamma $ matrices in each term. Using this, we have%
\begin{equation}
W\wedge W=\frac{1}{2}W_{A}\wedge W_{B}\left[ \Gamma ^{A},\Gamma ^{B}\right]
\ .  \label{W wedge W}
\end{equation}

A general identity of the $\Gamma $ matrices that we will use are%
\begin{eqnarray}
\Gamma ^{a_{1}\dots a_{p}}\Gamma _{bc} &=&\Gamma ^{a_{1}\dots
a_{p}}{}_{bc}-2p\Gamma ^{\lbrack a_{1}\dots a_{p-1}}{}_{[b}\delta
_{c]}^{a_{p}]}-\frac{p!}{\left( p-2\right) !}\Gamma ^{\lbrack a_{1}\dots
a_{p-2}}\delta _{\lbrack b}^{a_{p-1}}\delta _{c]}^{a_{p}]}\ , \\
\Gamma _{bc}\Gamma ^{a_{1}\dots a_{p}} &=&\Gamma _{ab}{}^{a_{1}\dots
a_{p}}-2p\delta _{\lbrack b}^{[a_{1}}\Gamma _{c]}^{a_{2}\dots a_{p}]}-\frac{%
p!}{\left( p-2\right) !}\delta _{\lbrack b}^{[a_{1}}\delta
_{c]}^{a_{2}}\Gamma ^{a_{3}\dots a_{p}]}\ .
\end{eqnarray}%

In particular, we can derive from it%
\begin{eqnarray}
\left[ \Gamma ^{ab},\Gamma _{c}\right]  &=&4\Gamma ^{\lbrack a}\delta
_{c}^{b]}\ ,\qquad \left[ \Gamma ^{ab},\Gamma _{cd}\right] =8\delta
_{\lbrack a}^{[c}\Gamma ^{d]}{}_{b]}\ , \\
\left[ \Gamma ^{a_{1}a_{2}a_{3}a_{4}a_{5}},\Gamma _{bc}\right]  &=&-20\Gamma
_{\lbrack b}{}^{[a_{1}a_{2}a_{3}a_{4}}\delta _{c]}^{a_{5}]}\ .  \notag
\end{eqnarray}%

Replacing (\ref{def W appendix}) into (\ref{W wedge W}) we get%
\begin{eqnarray}
W\wedge W &=&\frac{1}{2}\frac{1}{4}\omega _{ab}\wedge \frac{1}{4}\omega _{cd}%
\left[ \Gamma ^{ab},\Gamma ^{cd}\right] +\frac{1}{8^{2}}i\sigma _{2}\omega
_{ab}\wedge e^{c}\frac{1}{5!}F_{d_{1}\dots d_{5}}\left[ \Gamma ^{ab},\Gamma
^{d_{1}\dots d_{5}}\Gamma _{c}\right]   \notag \\
&&-\frac{1}{8^{3}}\left[ \slashed{F}_{5}\Gamma _{a},\slashed{F}_{5}\Gamma _{c}%
\right] e^{a}\wedge e^{c}\ .
\end{eqnarray}%

Using the commutator relations, we obtain%
\begin{eqnarray}
W\wedge W &=&\frac{1}{4}\omega _{ac}\wedge \omega _{\ b}^{c}\Gamma ^{ab}+%
\frac{1}{2}\frac{1}{8}i\sigma _{2}\omega _{\ b}^{a}\wedge e^{b}\slashed%
{F}_{5}\Gamma _{a}-\frac{1}{2}\frac{1}{8}i\sigma _{2}\omega ^{ab}\wedge e^{c}%
\frac{1}{4!}F_{d_{1}\dots d_{4}b}\Gamma ^{d_{1}\dots d_{4}}{}_{a}\Gamma _{c}
\notag \\
&&-\frac{1}{8^{3}}\left[ \slashed{F}_{5}\Gamma _{a},\slashed{F}_{5}\Gamma _{c}%
\right] e^{a}\wedge e^{c}\ .  \label{W wedge W to replace}
\end{eqnarray}%

Replacing (\ref{dW to replace}) and (\ref{W wedge W}) into (\ref{Xi in terms
of W}), the $2-$form integrability conditions become%
\begin{eqnarray}
\Xi  &=&\frac{1}{4}R_{ab}\Gamma ^{ab}+\frac{1}{16}i\sigma _{2}\frac{1}{5!}%
dF_{b_{1}\dots b_{5}}\Gamma ^{b_{1}\dots b_{5}}\Gamma _{a}e^{a}-\frac{1}{2}%
\frac{1}{8}i\sigma _{2}\omega ^{ab}\wedge e^{c}\frac{1}{4!}F_{d_{1}\dots
d_{4}b}\Gamma ^{d_{1}\dots d_{4}}{}_{a}\Gamma _{c}  \notag \\
&&-\frac{1}{8^{3}}\left[ \slashed{F}_{5}\Gamma _{a},\slashed{F}_{5}\Gamma _{c}%
\right] e^{a}\wedge e^{c}\ .
\end{eqnarray}%

Note that the second and third term form a Lorentz covariant derivative%
\begin{equation}
\Xi =\frac{1}{4}R_{ab}\Gamma ^{ab}+\frac{1}{16}i\sigma _{2}\frac{1}{5!}%
\mathcal{D}F_{b_{1}\dots b_{5}}\Gamma ^{b_{1}\dots b_{5}}\Gamma _{a}e^{a}-%
\frac{1}{8^{3}}\left[ \slashed{F}_{5}\Gamma _{a},\slashed{F}_{5}\Gamma _{c}\right]
e^{a}\wedge e^{c}\ .  \label{Xi simplificado}
\end{equation}%

The last term can be simplified by using $\left[ \Gamma ^{d_{1}\dots
d_{5}},\Gamma _{a}\right] =2\Gamma ^{d_{1}\dots d_{5}}{}_{a}$, then%
\begin{equation}
\left[ \slashed{F}_{5}\Gamma _{a},\slashed{F}_{5}\Gamma _{c}\right] e^{a}\wedge
e^{c}=4\slashed{F}_{5}\frac{1}{4!}F_{ab_{1}\dots b_{4}}\Gamma ^{b_{1}\dots
b_{4}}\Gamma _{c}e^{a}\wedge e^{c}-2\slashed{F}_{5}\slashed{F}_{5}\Gamma
_{ac}e^{a}\wedge e^{c}\ .  \label{comF5F5}
\end{equation}%

The last term of (\ref{comF5F5}) vanishes due to the fact that $F_{5}$ is
self-dual,%
\begin{eqnarray}
\left( 5!\right) ^{2}\slashed{F}_{5}\slashed{F}_{5} &=&F_{a_{1}\dots
a_{5}}F_{b_{1}\dots b_{5}}\Gamma ^{a_{1}\dots a_{5}}\Gamma ^{b_{1}\dots
b_{5}}\ , \\
&\sim &F^{d_{1}\dots d_{5}}\epsilon _{a_{1}\dots a_{5}d_{1}\dots
d_{5}}F_{c_{1}\dots c_{5}}\epsilon ^{c_{1}\dots c_{5}b_{1}\dots b_{5}}\Gamma
^{a_{1}\dots a_{5}}\Gamma _{b_{1}\dots b_{5}}\ ,  \notag \\
&=&F^{d_{1}\dots d_{5}}F_{c_{1}\dots c_{5}}\delta _{a_{1}\dots
a_{5}d_{1}\dots d_{5}}^{c_{1}\dots c_{5}b_{1}\dots b_{5}}\Gamma ^{a_{1}\dots
a_{5}}\Gamma _{b_{1}\dots b_{5}}\ ,  \notag \\
&=&F^{d_{1}\dots d_{5}}F^{a_{1}\dots a_{5}}\delta _{c_{1}\dots
c_{5}d_{1}\dots d_{5}}^{a_{1}\dots a_{5}b_{1}\dots b_{5}}\Gamma _{a_{1}\dots
a_{5}}\Gamma _{b_{1}\dots b_{5}}\;.  \notag
\end{eqnarray}%

Now we can anti-symmetrize and construct a $\Gamma _{a_{1}\dots a_{10}}$, and
then use the fact that it is proportional to $\epsilon _{a_{1}\dots
a_{10}}\Gamma _{11}$,%
\begin{eqnarray}
\left( 5!\right) ^{2}\slashed{F}_{5}\slashed{F}_{5} &=&F^{d_{1}\dots
d_{5}}F^{a_{1}\dots a_{5}}\delta _{c_{1}\dots c_{5}d_{1}\dots
d_{5}}^{a_{1}\dots a_{5}b_{1}\dots b_{5}}\Gamma _{a_{1}\dots a_{5}b_{1}\dots
b_{5}}\ , \\
&\sim &F^{d_{1}\dots d_{5}}F^{a_{1}\dots a_{5}}\Gamma _{d_{1}\dots
d_{5}a_{1}\dots a_{5}}\ ,  \notag \\
&\sim &F^{d_{1}\dots d_{5}}F^{a_{1}\dots a_{5}}\epsilon _{d_{1}\dots
d_{5}a_{1}\dots a_{5}}\Gamma _{11}\ .  \notag
\end{eqnarray}%

Note that the last line vanishes since it is equal to minus itself. Replacing
everything into (\ref{Xi simplificado}), we get the final form of the
integrability conditions%
\begin{equation}
\Xi =\frac{1}{4}R_{ab}\Gamma ^{ab}+\frac{1}{16}\frac{1}{5!}i\sigma _{2}%
\mathcal{D}F_{b_{1}\dots b_{5}}\Gamma ^{b_{1}\dots b_{5}}\Gamma _{a}e^{a}-%
\frac{1}{128}\frac{1}{4!}\slashed{F}_{5}F_{ab_{1}\dots b_{4}}\Gamma ^{b_{1}\dots
b_{4}}\Gamma _{c}e^{a}\wedge e^{c}\ .
\end{equation}

\section{Rotating D3-branes interpretation?}

We already saw that the 10-dimensional solution  (\ref{10dsol}) is understood as a deformation 
of a solution described by a continuous distribution of D3-branes. 

But we know \cite{Astefanesei:2007vh} that an extremal RNAdS solution (double Wick rotation of the 
RNAdS soliton), with constant scalars $X^i=X=$constant and 
equal gauge fields $A^i=A$ can be obtained as a limit from the 10-dimensional solution with angular momenta $l_i$, 
$i=1,2,3$ in 3 different (non-intersecting) planes,
\bea
&&ds^2=H^{-1/2}\left[-\left(1-\frac{2m}{r^4\Delta}\right)dt^2+dx_1^2+dx_2^2+dx_3^2\right]+H^{1/2}\left[
\frac{\Delta dr^2}{H_1H_2H_3-2m/r^4}\right.\nonumber\\
&&\left. +r^2\sum_{i=1}^3H_i(d\mu_i^2+\mu_i^2d\phi_i^2)
-\frac{4m\cosh \alpha}{
r^4H\Delta}dt \sum_{i=1}^3 \ell_i \mu_i^2d\phi_i+ \frac{2m}{r^4H \Delta}\left(\sum_{i=1}^3
\ell_i\mu_i^2d\phi_i\right)^2 
\right]\;,\label{rotd3}
\eea
where
\be
\Delta=H_1H_2H_3\sum_{i=1}^3\frac{\mu_i^2}{H_i};\;\;\;
H=1+\frac{2m\sinh^2\alpha}{r^4\Delta};\;\;\;
H_i=1+\frac{\ell_i^2}{r^2}.
\ee

So it is a reasonable question whether the current solution (\ref{10dsol}) cannot be obtained by a similar limit from the 
same. 

At first, things seem plausible. With 
\be
\mu_1=\cos\theta\sin\psi\;,\;\; \mu_2=\cos\theta\cos\psi\;,\;\;
\mu_1=\sin\theta\;,
\ee
and the rescaling (similar to, and inspired by the one in \cite{Astefanesei:2007vh})
\bea
m&=&\varepsilon^4 m'\;,\;\; \sinh\a =\varepsilon^{-2}\sinh\a'\;,\;\;\;
\ell_{1,2}=\varepsilon^2 \tilde \ell' \;,\;\;\; 
\ell_3=\varepsilon \ell'\;, \cr
r&=&\varepsilon r'\;,\;\; x^\mu=\varepsilon^{-1}x'^\mu\;,
\eea
followed by $\varepsilon\rightarrow 0$ and dropping the primes, one obtains 
\be
H_1=H_2=1\;,\;\; H_3=1+\frac{\ell^2}{r^2}=\left.\lambda^6\right|_{\epsilon=+1}\;,\;\;
\Delta=1+\frac{\ell^2}{r^2}\cos\theta=\left.\zeta^2\right|_{\epsilon=+1}\;,
\ee
and so the coefficient of $d\vec{x}_{1,2}^2$ matches,
\be
H^{-1/2}d\vec{x}_{1,2}\rightarrow \left(\frac{2m\sinh^2\a}{r^4 \zeta^2}\right)^{-1/2}=\frac{\zeta r^2}{L^2}\;,\;\;
L^4\equiv 2m\sinh^2\a>0\;,
\ee
and one finds also matching for the coefficients of $d\phi_1^2,d\phi_2^2,d\phi_3^2$, which are 
$H^{1/2}r^2H_i \mu_i^2$ (note that $\frac{2m}{r^4H\Delta}\ell_i^2\sim \varepsilon^6$ is subleading in $\varepsilon$ with respect to 
$r^2H_i^2\sim \varepsilon^2$, so is dropped), and of $\sum_i H_i d\mu_i^2=\zeta^2d\theta+\cos^2\theta d\psi^2$, 
which is $r^2 H^{1/2}=L^2/\zeta$. 

The problem comes in the interpretation of the terms with $A_i$ and $d\phi$, and of the $dr^2$ term. Matching of the 
$dr^2$ coefficient results in the equation 
\be
2m=\ell^2L^2\left[q_1^2\left(1+\frac{\ell^2}{r^2}\right)-q_2^2\right]\Rightarrow \frac{\ell^2}{L^2}(q_1^2-q_2^2)\simeq
\frac{1}{\sinh^2\a}\;\;\;{\rm for}\;\; r\gg \ell\;,
\ee
which could only be satisfied approximately, for $r\gg \ell$ and $q_2<q_1$, due to the $1/r^2$ term (note that $q_1=0$
does not work, since it implies $m<0$). 

Matching of the terms with $A_id\phi_i$, after the double Wick rotation, replacing $dt$ from the rotating D3-brane solution 
with the $d\phi$ from the soliton solution, is only possible in some approximate sense as well,
but now also with $r-r_0\sim \varepsilon$ or $\sim \varepsilon^2$ fixed, since in the soliton $d\phi_i
A_i$ is proportional to $q_1\frac{\ell^2}{L}\frac{r^2-r_0^2}{r_0^2}$ or $q_2\frac{\ell^2}{L}\frac{r^2-r_0^2}{r_0^2+4\ell^2}$, 
while the former has (at least) an extra power of $\varepsilon$, and so is proportional to $(\varepsilon^2\tilde\ell) 4m \cosh \a\simeq 
(\varepsilon^2\tilde \ell)2L^4/\sinh \a$ or $(\varepsilon\ell) 4m \cosh\a\simeq (\varepsilon\ell) 2L^4 /\sinh \a$, respectively, 
so one would have to consider some unusual simultaneous near-horizon limit, depending on the charge. 

Moreover then, the coefficient of the $d\phi^2$ term, composed of $(\zeta r^2/L^2) F(r)L^2=H^{-1/2}F(r)L^2$ 
and the $H_i\mu_i^2A_i^2$ terms, would have to match $H^{-1/2}\left(1-2m/r^4\Delta\right)=H^{-1/2}\left(1-2m/(r^4\zeta^2)\right)$,
which depends on the previous near-horizon limit.

In conclusion, the deformation found in this paper is a nontrivial deformation of the rotating D3-brane solution, that is 
not easily understandable within the same context, except maybe in some generalized near-horizon sense.

\newpage

\hypersetup{linkcolor=blue}
\phantomsection 
\addtocontents{toc}{\protect\addvspace{4.5pt}}
\bibliographystyle{mybibstyle}
\bibliography{bibliografiaD3solitons}

\end{document}